\documentclass[pre,a4paper,twocolumn,superscriptaddress,%
floatfix]{revtex4}

\usepackage[T1]{fontenc}
\usepackage[utf8]{inputenc}

\usepackage{graphicx}
\usepackage{color}
\usepackage{amsmath}

\begin{document}

\title{Memory effects in glasses: insights into the thermodynamics of out of
  equilibrium systems revealed by a simple model of the Kovacs effect }

\author{Michel Peyrard}
\affiliation{Universit{\'e} de Lyon, Ecole Normale Sup{\'e}rieure de Lyon,
Laboratoire de Physique CNRS UMR 5672, 46 all{\'e}e d'Italie, F-69364 Lyon
Cedex 7, France}

\author{Jean-Luc Garden}
\affiliation{Univ. Grenoble Alpes, CNRS, Grenoble INP, Institut N\'{E}EL,
38000 Grenoble, France}

\date{\today}

\begin{abstract}
{\em This paper is an extended version of an article accepted for publication
  in Physical Review E. Besides its fundamental interest, the model that we
  investigate in this article is simple enough to be used as a basis for
  courses or 
  tutorials on the thermodynamics of out of equilibrium systems. It allows
  simple numerical calculations and analytical analysis which highlight
  important concepts with an easily workable example. This version includes
  studies of fast cooling and heating, exhibiting cases with negative
  heat capacity, and further discussions on the entropy
  which are not presented in the Physical Review E version.}

~

Glasses are interesting materials because they allow us to explore the
puzzling properties of out-of-equilibrium systems. One of them is the Kovacs
effect in which a glass, brought to an out-of-equilibrium state in which all
its thermodynamic variables are identical to those of an equilibrium state,
nevertheless evolves, showing a hump in some global variable before the
thermodynamic variables come back to their starting point. We show that a
simple three-state system is sufficient to study this phenomenon using
numerical integrations and exact analytical calculations.
It also brings some light on
the concept of fictive temperature, often used to extend standard
thermodynamics to the out-of-equilibrium properties of glasses. We confirm that
the concept of a unique fictive temperature is not valid, an show it can be
extended to make a connection with the various relaxation processes in the
system. The model also brings further insights on the thermodynamics of
out-of-equilibrium systems.
Moreover we show that the three-state model is able to
describe various effects observed in glasses such as the asymmetric
relaxation to equilibrium discussed by Kovacs, or the reverse crossover
measured on $B_2O_3$.
\end{abstract}

\maketitle

\section{Introduction}

Glasses are very common materials, and nevertheless they are very special,
with fascinating properties which pose fundamental questions to
physicists. When a glass is formed by cooling a liquid, the dynamics of
molecular reorganization drastically slows down in a narrow temperature range,
so that, below a temperature $T_g$, called the glass-transition temperature,
or above some pressure $P_g$, the fluctuations within the material appear as
frozen on observable time scales \cite{KOVACS}. The mechanisms behind this
transition are only partly understood \cite{BERTHIER2011}, but it is not the
only question that glasses pose to theoreticians because, below $T_g$, they are
out-of-equilibrium systems. Standard thermodynamics does not apply and this
opens an entirely new world, with many puzzling properties. For instance
experiments show that the heat capacity of a glass depends on its past history
\cite{THOMAS1931}.

The Kovacs effects is another example which provides a revealing insight into
the properties of out-of-equilibrium systems. Let us consider a glassy
material, in the $V$, $T$ thermodynamic representation, that started from an
initial equilibrium at temperature $T_0$ and was slowly cooled until it
reaches a new equilibrium at a temperature $T_1$ with a specific volume
$v(T_1)$. Let us call this state state $A$.
Then, in a second experiment the same piece
of material is abruptly cooled from $T_0$ to a temperature
$T_2 < T_1$, $T_2$ being in
the range of the glass transition temperature $T_g$. The material is let to
age at $T_2$. Its specific volume is monitored
while it slowly decreases with aging.
When the specific volume reaches the value $v(T_1)$ the material
is quickly brought to temperature $T_1$. This new state, that we call state
$B$, has now the same thermodynamic variables as state $A$, and we would
therefore expect state $B$ to be in equilibrium. However, when state $B$ was
kept at temperature $T_1$, Kovacs observed that its specific volume started to
increase, then passed a maximum and decayed to reach the value $v(T_1)$
again.
Obviously state $B$ was not the same as state $A$ although both had the same
thermodynamic equilibrium variables. This surprising experiment shows that
{\em the thermodynamic variables are not
sufficient to characterize the state of out-of-equilibrium system such as a
glass.} There are further ``hidden variables'' which must also be
specified.
This idea brings two kinds of questions: i) what are the system features which
can lead to such a situation, ii) what are its consequences and how do
they affect the thermodynamic description of the system. The Kovacs effect has
been widely studied with a particular attention on the first question
\cite{BERTHIER2002,BERTIN,AQUINO,MOSSA}. These studies consider various models
which can exhibit glassy properties, with a distribution of length or time
scales, and the coupling of many degrees of freedom, which can lead to memory
effects. The second class of questions has been less explored. An article by
Bouchbinder and Langer \cite{BOUCHBINDER} examines the non-equilibrium
thermodynamics of the Kovacs effect and lists the minimal ingredients for any
model of this effect. However the view point that complexity is inherent to
this phenomenon is suggested by the introductory sentence: ``The Kovacs effect
reveals some of the most subtle and important nonequilibrium features of
glassy dynamics''. Actually, as we show here, the Kovacs effect exists in an
extremely simple system, so simple that it may appear almost trivial. In spite
of its simplicity, and
perhaps because it is so simple, this model deserves a study because it allows
a complete understanding of the mechanisms behind the Kovacs effect and brings
a new light on the concept of fictive temperature \cite{DAVIESJONES} widely
used for glasses. It is now recognized that memory effects are not compatible
with a unique fictive temperature \cite{LEUZZI}. Our analysis confirms this
view but moreover shows quantitatively why, and how the concept can be
extended and generalized.

\medskip
{\em Besides its fundamental interest, the model that we
  investigate in this article is simple enough to be used as a basis for
  courses or 
  tutorials on the thermodynamics of out of equilibrium systems. It allows
  simple numerical calculations and analytical analysis which highlight
  important concepts with an easily workable example. }

\medskip
Section \ref{sec:model} introduces the model, and discusses its relationship
with an even simpler model which was used to illustrate non-equilibrium
negative heat capacity in glasses. Section \ref{sec:numerical} shows how
simple numerical integrations of this model exhibit the Kovacs effect.
An analytical description is used in
Sec.~\ref{sec:analytical} to recover the numerical results while providing a
deeper understanding of the mechanisms behind these results and other
unusual properties of glasses. Finally
Sec.~\ref{sec:fictiveT} examines to what extend these properties can be
described by an extension of standard thermodynamics using the concept
of ``fictive temperature'' \cite{DAVIESJONES}, and how the out-of-equilibrium
properties  appear in the entropy, and entropy
production, for the different processes studied with this model.

\section{Model}
\label{sec:model}

The glass transition can be observed for a large variety of systems, from
liquids bonded by covalent, ionic or molecular forces, to granular
materials or proteins. Therefore its microscopic description is not unique,
but there is a unifying concept which underlies the various phenomena which
are involved, the free energy landscape over which the system evolves. This
highly multidimensional surface is itself very complex, but many properties of
glasses can be derived from a reduced view, the inherent structure landscape
which only describes the minima of the metastable states \cite{NAKAGAWA}.
This is enough to build a thermodynamics of the equilibrium properties of a
glass, and, if the picture is completed by adding the barriers that have
to be overcome to move between the minima, its dynamical properties can also
be investigated. This
idea can be pushed to the extreme by considering only a small number of energy
states and the saddle points which separate one state from another.

\begin{figure}[h]
  \centering
  \includegraphics[width=7cm]{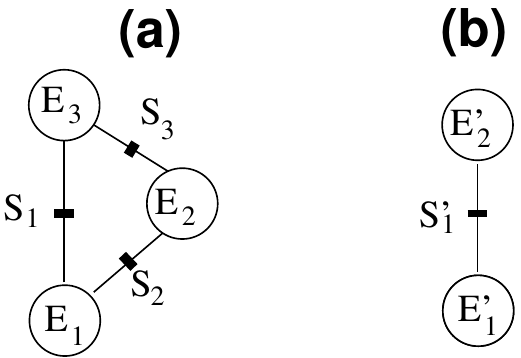}
  \caption{Picture of the three-state model (a) and its simplified
    version, the two-state model (b). The circles schematize the metastable
    states with energies $E_i$, and the thick lines the barriers that separate
    them, with energies $S_i$. }
  \label{fig:model}
\end{figure}

\medskip
The simplest model includes only two energy states, separated by one saddle
point. This two-level system is sufficient to study the negative heat capacity
observed in the vicinity of the glass transition, but, as shown below, it is
nevertheless unable to describe other properties of glasses, such as the
Kovacs effect. However, adding only one energy state, and two saddle points,
as schematized on Fig.~\ref{fig:model}-a, is enough to illustrate fundamental
ideas of the physics of out-of-equilibrium systems.

Let us denote by $E_i$ ($i = 1,2,3$) the energies of the three metastable
states.  The probabilities $P_i$ that the states $i$ are occupied are the
variables which define the state of the system. However, due to the constraint
$P_1 + P_2 + P_3 = 1$, the state is actually defined by two parameters
only. We henceforth select the two variables $P_1$ and $P_2$ to characterize a
state of the system.

\medskip
The equilibrium properties of the model are readily obtained from the
Gibbs canonical distribution. Its partition function is 
\begin{equation}
  \label{eq:z}
  Z = \sum_i e^{-E_i/T} \; ,
\end{equation}
if we measure the temperature $T$ in energy units (which is equivalent to
setting the Boltzmann constant to $k_B = 1$). The occupation probabilities of
the three states when the system is in equilibrium are 
\begin{equation}
  \label{eq:pieq}
  P_i^{\mathrm{eq}} = \frac{1}{Z} e^{-E_i/T} \; ,
\end{equation}
and the average energy of the system is
\begin{equation}
  \label{eq:emoyeq}
  E^{\mathrm{eq}}(T) =
  \langle E(T) \rangle = \frac{1}{Z} \; \sum_i E_i e^{-E_i/T} \; .
\end{equation}

\medskip

Viewing
this model as a simplified picture of the free energy landscape of a glass, we
assume that the transitions from one basin of attraction to another are
thermally 
activated over saddle points having energies $S_1$, for the transition between
$E_1$ and $E_3$, $S_2$ for the transition between $E_1$ and $E_2$, and $S_3$
for the transition between $E_2$ and $E_3$. 
Therefore the transition
probabilities are determined by a  set barriers $B_{i,j}$,
for instance 
$B_{1,3} = S_1 - E_1$, $B_{3,1} = S_1 - E_3$, $B_{1,2} = S_2 - E_1$, and so on.
The energies of the saddle points
are assumed to be higher than the energies of the states that they separate so
that $B_{i,j} >0$ for all $i,j$ pairs.

The rates of the thermally activated transitions are
\begin{equation}
  \label{eq:wij}
  W_{i \to j} = \omega_{ij} \; e^{- B_{ij}/T}
\end{equation}
where $\omega_{ij}$ are model parameters which have the dimension of inverse
time. As a result the thermodynamics of the model is expressed by equations
for the time-dependence of the occupation probabilities, which are of the form
\begin{align}
  \label{eq:dp1dt}
  \frac{d P_1}{dt} = &- P_1 \, \omega_{13} \, e^{-(S_1 - E_1)/T} +
                       P_3 \, \omega_{31} \, e^{-(S_1 - E_3)/T}  \nonumber \\
                     & - P_1 \, \omega_{12} \, e^{-(S_2 - E_1)/T} +
                  P_2 \, \omega_{21} \, e^{-(S_2 - E_2)/T} \; ,
\end{align}
and similar equations for $P_2$ and $P_3$.

For the equilibrium probabilities $P_i^{\mathrm{eq}}$,
Eq.~(\ref{eq:dp1dt}) becomes
\begin{align}
  \left(\frac{d P_1}{dt}\right)_{\mathrm{eq}}
  = &- \frac{1}{Z}e^{-E_1/T} \, \omega_{13}
                       \, e^{-(S_1 - E_1)/T} \nonumber \\ &+
                       \frac{1}{Z}e^{-E_3/T} \, \omega_{31}
                       \, e^{-(S_1 - E_3)/T}  \nonumber \\
                     & -  \frac{1}{Z}e^{-E_1/T} \, \omega_{12}
                       \, e^{-(S_2 - E_1)/T} \nonumber \\ &+
                       \frac{1}{Z}e^{-E_2/T} \, \omega_{21}
                       \, e^{-(S_2 - E_2)/T} \; .
\end{align}
If we assume that $\omega_{13} = \omega_{31}$ the first two terms cancel each
other, and if we assume that $\omega_{12} = \omega_{21}$, terms 3 and 4 cancel
each other so that $(dP_1/dt)_{\mathrm{eq}}$
vanishes, as we expect for the equilibrium case.
These conditions $\omega_{ij} = \omega_{ji}$ are the so-called ``detailed
balance'' conditions, which ensure the existence of the equilibrium
state. Detailed balance does not require $\omega_{13} = \omega_{12}$, however
we shall henceforth assume
\begin{equation}
  \label{eq:omeg1}
  \omega_{ij} = 1 \qquad \forall i, j (i \not= j) \;.
\end{equation}
Setting the common value of $\omega_{i,j}$ to $1$ defines the time unit (t.u.)
for the system.

\medskip

This model belongs to a class of systems described by a master equation which
have been investigated in \cite{PradosJSTAT} in a study which derived general
properties of the Kovacs hump valid for a variety of systems. Instead our goal
here is to show that a minimal model, for which an exact analytical analysis
is straightforward, is able not only to generate the Kovacs
effect but also other properties observed in real materials or complex
models, as discussed in Sec.~\ref{sec:discussion}. Moreover we are also
interested in the thermodynamics of out equilibrium systems (Sec.
\ref{sec:fictiveT}). This model allows
us to generalize the concept of fictive temperature,
introduced in 1931 \cite{TOOL1931} and still widely used for glasses
\cite{WONDRACZEK,GUO-2011} but which fails for systems exhibiting
the Kovacs effect.

\section{Numerical studies}
\label{sec:numerical}

Numerical integrations of Eq.~(\ref{eq:dp1dt}), taking into account our choice
(\ref{eq:omeg1}) for $\omega_{i,j}$, provide a first insight of the properties
of the model. We selected the following parameters. The energy level $E_3$ has
been used as the reference level for the energies by setting $E_3 = 0$. The
energies of the two other states were chosen as $E_1=-0.40$ and
$E_2=-0.25$. The energies of the saddle points were set to $S_1 = 0.40$, $S_2
= 0.30$, $S_3 = 0.25$. As the model does not intend to describe a specific
glass, the energy scale is irrelevant and the values have been chosen
arbitrarily. Other choices would of course quantitatively modify the results
but, as long as the conditions $B_{ij}>0$ are verified and the ratios of the
investigated temperatures and energies stay in the same range, the main
features of the results would be preserved.

\begin{figure}[h]
  \centering
  \includegraphics[width=8cm]{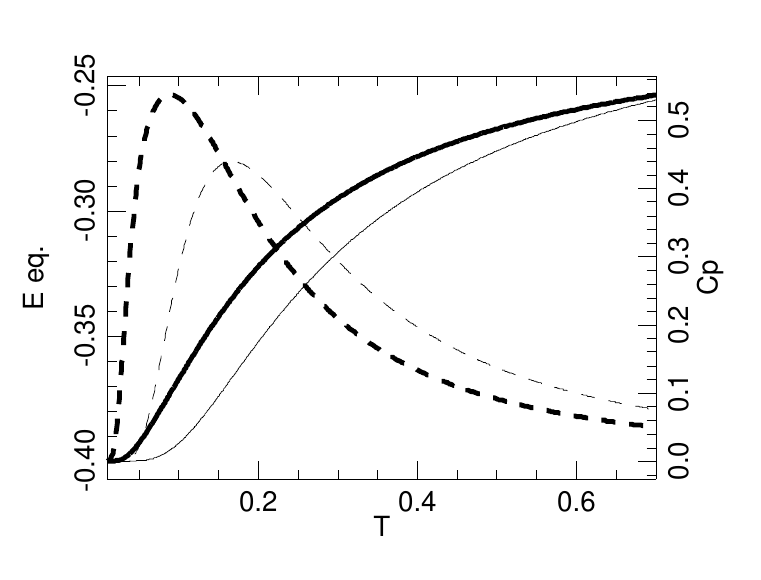}
  \caption{Equilibrium properties of the model: energy (full lines) and
    heat capacity (dashed lines) versus temperature. The thick lines show the
    results for the three-state model, and the thin lines show the results for
  the two-state model for comparison.}
  \label{fig:eqECp}
\end{figure}

Figure \ref{fig:eqECp} shows the equilibrium energy and heat capacity versus
temperature for the three-state model of Fig.~\ref{fig:model}-a. For
comparison it also shows the same quantities for the two-state model
of  Fig.~\ref{fig:model}-b, with the parameters $E'_1 = -0.40$, $E'_2=0$, and
$S'_1 = 0.40$ chosen so that this restricted model simply corresponds to the
three-state model without the intermediate state $E_2$. The two models have
qualitatively similar equilibrium properties.

\bigskip
In a typical experiment, the system is in equilibrium at a temperature
$T_{\mathrm{ini}}$ at time $t=0$. Then its temperature is varied by choosing a
series of set-points $T^s_j$ and defining the time points $t_j$ at which each
set-point has to be reached. This defines a trajectory $T(t)$ which is imposed
to the system while its state variables $P_i(t)$ are followed. The evolution
of the system is defined by Eq.~(\ref{eq:dp1dt}) and the corresponding
equations for $P_2(t)$ and $P_3(t)$, with constrained values for $T(t)$ in the
right hand sides. The initial values of the state variables are the values
$P_i^{\mathrm{eq}}(T_{\mathrm{ini}})$ which are known since they are given by
Eq.~(\ref{eq:pieq}). Any standard algorithm can be used to solve the set of
three coupled equations for $P_1$, $P_2$, $P_3$. In our calculations we used
the $4^{\mathrm{th}}$-order Runge-Kutta method \cite{CARNAHAN}. The time step
$\delta t$ can be chosen according to the characteristic relaxation times of
the system (see. Sec. \ref{sec:analytical}), however, as the integration only
involves the solution of three coupled differential equations, choosing a small
value such as $\delta t = 0.01$ ensures a high accuracy without a significant
computational cost. As the system only has two independent variables, $P_1$
and $P_2$, the computation could be simplified by solving only two coupled
equations, completed by the condition $P_3 = 1 - (P_1 + P_2)$. However,
solving the three coupled equations is interesting because it allows a simple
test of accuracy, and a check of the value of $\delta t$, by making sure that
the condition $\sum P_i = 1$ is well preserved by the calculation.

\medskip


\subsection{Fast cooling and heating}

As a first example of the properties of the three-state model, let us consider
a first simulated experiment involving a simple cooling at fixed cooling rate,
followed by heating at the same rate. The temperature set-points are
$T_{\mathrm{ini}} = 0.20$, $T^s_1 = 0.02$ at time $t_1 = 10^4\;$t.u. ,
$T^s_2=0.20$ at $t_2 = 2\; 10^4\;$t.u.~.
\begin{figure}[h]
  \centering
  \begin{tabular}{c}
    \\
    (a) \\
    \includegraphics[width=8cm]{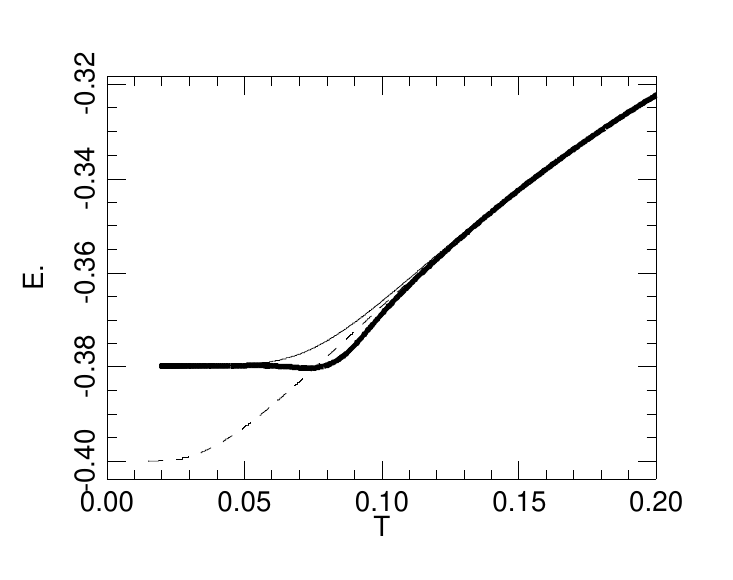} \\
    (b) \\
    \\
    \includegraphics[width=8cm]{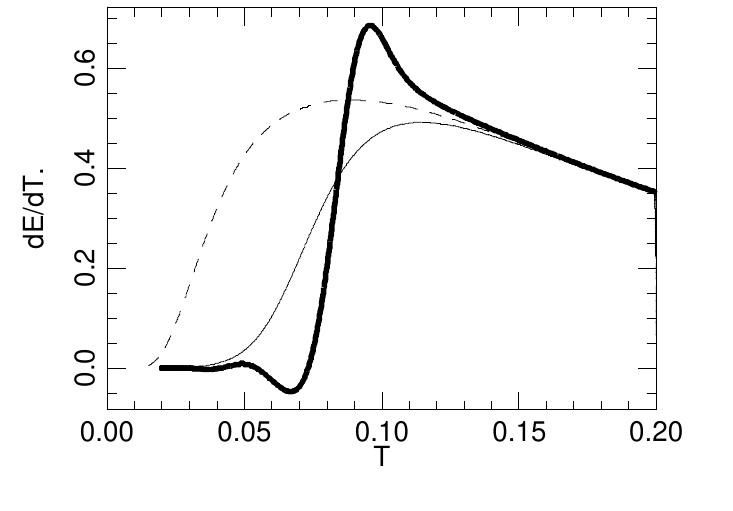}
  \end{tabular}
  \caption{Energy (a) and heat capacity (b) versus temperature when the
    three-state system is cooled from $T_{\mathrm{ini}} = 0.20$ to
    $T^s_1 = 0.02$ in $t_1 = 10^4$ time units, with a constant cooling rate,
    and then heated again from the
    final state back to $T^s_2=0.20$ in the same time interval. The thin full
    lines show the properties of the model upon cooling, while the full thick
    black lines show its behavior upon heating. The dashed lines show the
    equilibrium properties of the model versus temperature.
  }
  \label{fig:coolheat}
\end{figure}

Figure \ref{fig:coolheat}-a shows the variation of the total energy of the
system versus temperature. On cooling, it shows that, above $T
\approx 0.1$ the energy (thin full line)
follows the curve of the equilibrium energy versus $T$ (dashed line).
But when the
temperature decreases further, the energy starts to decay more slowly than the
equilibrium energy and stays almost constant when $T < 0.05$. The temperature
range $0.05 < T < 0.10$ is the domain of the glass transition, below which the
dynamics of the system tends to freeze. The large deviation between the
observed value of the energy of the system after cooling and the equilibrium
energy at the same temperature, shows that, with the cooling rate that we used
for this simulation, the system was brought to a strongly out-of-equilibrium
state. If the simulation is repeated with a larger value of $t_1$, i.e. with a
smaller cooling rate, the energy follows the equilibrium energy down to lower
temperatures, as expected since we gave the system a longer time to approach
equilibrium.

\smallskip
When the system is heated from the
state that it reached after cooling, the energy (full thick curve)
stays constant and {\em even
slightly decreases} in the temperature range of the glass transition and
passes below the curve of the equilibrium energy versus temperature, before
rising sharply to join the curve of the equilibrium energy. Figure 
\ref{fig:coolheat}-b, showing the heat capacity $C = dE/dT$ emphasizes this
behavior. The heat capacity versus temperature
has the characteristic shape
observed when a glassy material is heated after a rapid cooling
\cite{THOMAS1931}, with a hump after a sharp rise before it reaches the
equilibrium heat capacity at high temperature.
In this case, as cooling was 
fast enough to bring the system very far from equilibrium, the fast rise is
preceded by domains in which the
heat capacity is {\em negative}, a
phenomenon often observed in calorimetric heating scans through the glass
transition \cite{DEBOLT}.
This simulation shows that the three-state system can exhibit the typical
properties of a glassy system. This kind of behavior was also observed for the
simpler two-state system of Fig.~\ref{fig:model}-b
\cite{BISQUERT,JABRAOUI}.
However, 
to explore more subtle glassy properties, such as the Kovacs effect, the three
states are necessary.

\subsection*{Kovacs effect}

\begin{figure}[h]
  \centering
  \begin{tabular}{c}
    (a) \\
    \includegraphics[width=8cm]{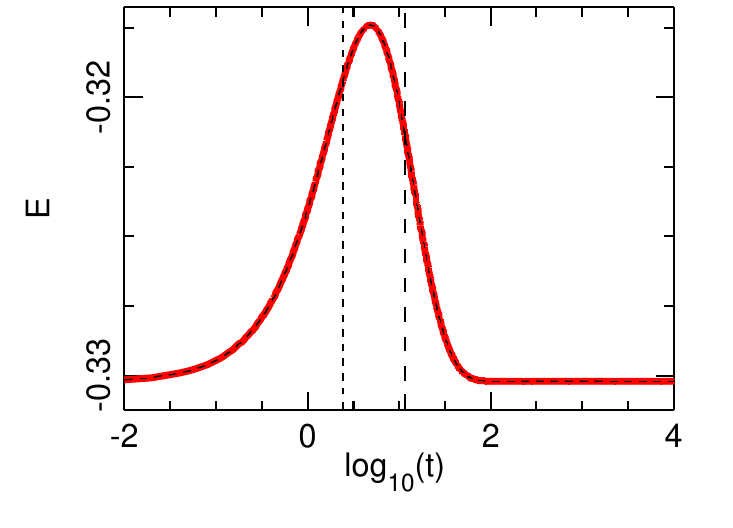} \\
    (b) \\
    \includegraphics[width=8cm]{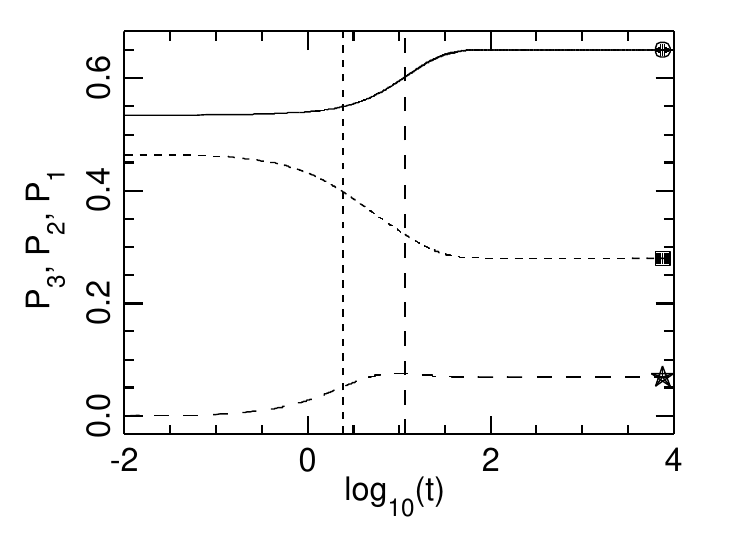}
  \end{tabular}
  \caption{Energy (a) and probabilities $P_1$,$P_2$, $P_3$ (b) versus
    time (in log scale) during the time evolution
    at constant temperature $T_2 = 0.17856$ after aging for $1.8\,10^7\,$t.u.
    at temperature $T_1 = 0.02$ in the study of the Kovacs effects.
    In plot (a), the thick red line shows the numerical result and the
    short-dash line shows the result of the theoretical calculation
    (Sec.~\ref{sec:analytical}). 
    Plot (b): full line $P_1$, short-dash line $P_2$, long-dash line $P_3$.
    The points at the right side of the graph show the values of the
    equilibrium probabilities at $T_2$ : circle $P_1^{\mathrm{eq}}$,
    square $P_2^{\mathrm{eq}}$, star $P_3^{\mathrm{eq}}$.
    In both plots, the vertical lines mark the relaxation times
    (short-dash line $\tau_1$, long-dash line $\tau_2$).
  }
  \label{fig:kovacs1}
\end{figure}

The Kovacs effect is observed when one
follows some property, such as the specific volume, of a glassy material
subjected to a particular protocol. For the three-state system the volume has
no meaning but we can follow another variable, such as the energy $E = \sum
P_i E_i$. We do not need to perform the preliminary scan with a slow cooling
mentioned in the introduction because we know the equilibrium energy
$E^{\mathrm{eq}}(T)$ from Eq.~(\ref{eq:emoyeq}). Let us consider the following
thermal protocol $T(t)$: we start from an equilibrium state at
$T^{\mathrm{ini}} = 0.30$. At time $t=0$ the system is abruptly cooled to $T_1
= 0.02$ and we let it age at this temperature until $t_1 = 1.8\,10^7$t.u.~. At
the end of this aging period its energy has decreased to a value $E_1$. We
determine the temperature $T_2$ at which this energy would be the equilibrium
energy by solving the equation $E^{\mathrm{eq}}(T_2) = E_1$, which gives
$T_2=0.17856$. At this instant $t_1$, the temperature of the system is brought
to $T_2$ and we follow its evolution for $10^4$t.u.~. As its energy
corresponds to the equilibrium energy $E^{\mathrm{eq}}(T_2)$, we 
could expect that nothing happens and that the system simply stays in
equilibrium. As shown by Fig.~\ref{fig:kovacs1}-a, this is not what is observed.

The energy starts to grow, shows a hump, and decays back to its original
value. This is exactly what Kovacs observed for the volume of a vinyl
polyacetate sample, with a similar protocol \cite{KOVACS}.
Figure~\ref{fig:kovacs1}-b helps to understand what happens. It shows that, at
the end of the aging period, when temperature was abruptly raised to $T_2$,
the occupation probabilities $P_i$ of the tree states were very different from
the equilibrium probabilities  $P_i^{\mathrm{eq}}(T_2)$. Although the energy
of the system was $E_1=E^{\mathrm{eq}}(T_2)$, the system was {\em not} in the
equilibrium state at $T_2$. This is possible because the three-state system has
{\em two} internal degrees of freedom $P_1$, $P_2$. A given value of its
energy $E = P_1 E_1 + P_2 E_2 + [1 - (P_1+P_2)] E_3$ can be realized by
different microscopic states, characterized by $P_1$ and $P_2$. If the system
is prepared in a non-equilibrium state by a specific protocol (a temperature
quench followed by aging at low temperature in Kovacs experiment) then the
internal degrees of freedom do not have their equilibrium values. In the
time evolution at temperature $T_2$ 
the internal variables evolve towards equilibrium. As 
shown by the theoretical analysis of Sec.~\ref{sec:analytical}, this process
is a relaxation, which does not guarantee that $E$ should stay constant. And
actually it does not, and we observe the hump characteristic of the Kovacs
effect.

\medskip
The scheme that we described above appears to be very general. A macroscopic
system has many internal degrees of freedom, and therefore many internal
configurations which could lead to the same energy (or the same specific
volume, for the variable studied in Kovacs experiment). So, why don't we
observe the Kovacs effects for almost any material. The answer lies in the
possibility to prepare the system in an adequate
out-of-equilibrium state. Usually the
time scale at which the internal variables approach equilibrium is too fast to
allow us to observe the Kovacs effect. This is different for glasses which are
material in which some degrees of freedom are almost frozen or at least evolve
so slowly that we can follow their evolution. In the experiment of Kovacs, the
characteristic time of this evolution was of the order of several hundreds
hours.

\medskip
What we have described here for the three-state system cannot be observed for
the two-state system. If a calculation is carried with the same protocol for
the two-state system, when the temperature is kept constant at $T_2$ after
aging, the energy stays strictly constant. The Kovacs hump is not observed
with this model. It is easy to understand why. The two state-system only has
one degree of freedom. Its energy is $E = P_1 E_1 + [1 - P_1] E_2$.
There is a one-to-one correspondence between $E$ and $P_1$.
Therefore looking for $T_2$ by solving $E_1 = E^{\mathrm{eq}}(T_2)$ is
equivalent to solving $P_1 = P_1^{\mathrm{eq}}(T_2)$, i.e., when we start
the time evolution at $T_2$, the system is already in its equilibrium state.
Some studies have reported the Kovacs effects in models based
on two-state systems \cite{PradosJSTAT,AQUINO2008}, but these models also
include disorder, either through dynamic fluctuations of the height of the
barrier between the states, or through a distribution of model parameters.
This extends the configuration space of the model so that the energy of the
system does not fully determine the configuration of the system.

\begin{figure}[h]
  \centering
  \includegraphics[width=8cm]{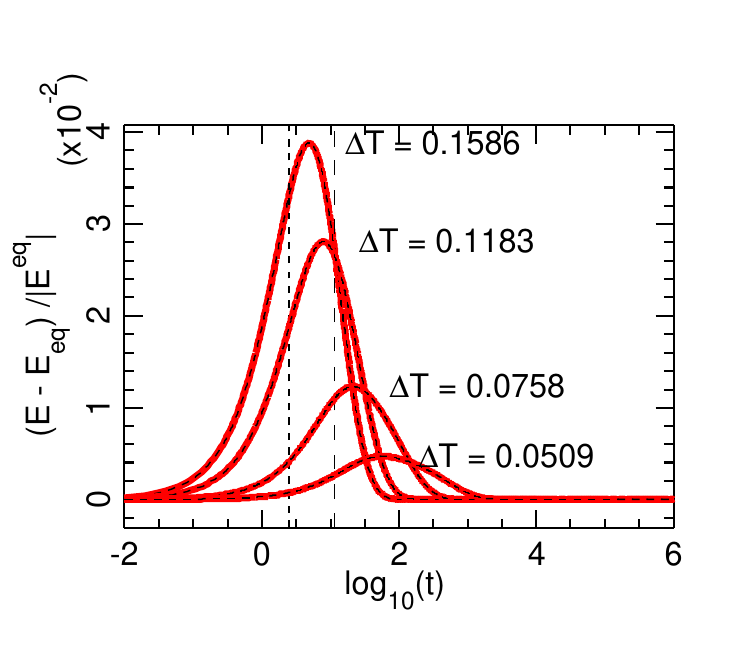}
  \caption{Normalized
    Kovacs hump $(E - E^{\mathrm{eq}})/|E^{\mathrm{eq}}|$
    versus time (in log scale) for various values of the temperature jump
    $\Delta T = T_2-T_1$. The thick red lines are the results of the numerical
    integrations and the black dashed lines are analytical calculations
    (Sec.~\ref{sec:analytical}). } 
  \label{fig:kovacs2}
\end{figure}

\medskip
To make sure that we do observe the Kovacs effect, it is necessary to check
that the numerical results obtained with the three-state system match the
various experimental features of this effect. Besides the existence of the
Kovacs humps, experiments show that the hump gets smaller,
and appears at a later
time when the difference between the temperature $T_1$ of the
aging period and that of the Kovacs time evolution at $T_2$ decreases.
Starting from the same initial
equilibrium at $T^{\mathrm{ini}}=0.30$ we have performed calculations for
different values of the aging temperature $T_1$, $T_1 = 0.030,\; 0.032,\; 0.033$
with the same aging time
$t_1 = 1.8\,10^7$t.u.~. The final energy $E_1$ depends on $T_1$ and
therefore the temperature $T_2$ such that $E_1= E^{\mathrm{eq}}(T_2)$ also
changes. Figure \ref{fig:kovacs2} shows that the hump depends on
$\Delta T =T_2 - T_1$ as observed by Kovacs \cite{KOVACS}. For the
three-state system this behavior can be justified analytically as discussed in
Sec.~\ref{sec:analytical}. 

\section{Analytical analysis}
\label{sec:analytical}

A configuration of the three-state system depends on the two variables $P_1$
and $P_2$. Their variation versus time could be obtained by numerical
integrations, but, as Eq.~(\ref{eq:dp1dt}) and the corresponding equation for
$P_2$ are a set of coupled first order linear differential equations, they can
also be solved analytically.

It is convenient to introduce the deviations with respect to equilibrium
$Q_i = P_i - P_i^{eq}$. The condition $P_1 + P_2 + P_3 = 1$ results in
$Q_1 + Q_2 + Q_3 = 0$.
Note that there is no assumption regarding the size of $Q_i$ compared to
$P_i^{eq}$, i.e.\ this change of variable {\em does not imply any small
amplitude expansion}. Actually when the system is subjected to an abrupt
temperature change, immediately after the temperature jump the $P_i$
have preserved their values while the $P_i^{eq}$ have drastically
changed. This may lead to situations where $|Q_i/P_i| \gg 1$.

As shown in Appendix \ref{app-solution}, the solution of the system of linear
equations for $Q_1$, $Q_2$ can be expressed in terms of two eigenmodes as
\begin{equation}
    \label{eq:decomposQ}
    \vec{Q} = a(t) \vec{U}^{(1)} + b(t) \vec{U}^{(2)} \; .
  \end{equation}
where $\vec{U}^{(1)}$ and $\vec{U}^{(2)}$ are two eigenvectors and $a(t)$,
$b(t)$, the amplitudes of the modes, are given by two exponential relaxations
\begin{align}
  \label{eq:solab}
  a(t) &= a(t=0) \exp[ - t / \tau_1] \nonumber \\
  b(t) &= b(t=0) \exp[ - t / \tau_2] \; ,
\end{align}
characterized by two relaxation times $\tau_1$, $\tau_2$.
Both the eigenvectors and the relaxation times depend on the parameters of the
model and on temperature. For any constant-temperature process the modes are
well defined and the time dependencies of their amplitudes are readily obtained
from their values at the beginning of the evolution period, which are
themselves determined from the initial values of the probabilities $P_i(t=0)$
through the definition of the $Q_i$s and Eq.~(\ref{eq:decomposQ}). For a
protocol starting from an equilibrium state, and comprising only temperature
jumps, during which the probabilities $P_i$ do not change, and time evolutions
at constant temperature this provides a systematic procedure to derive an
exact analytical solution which determines the complete evolution of the
system.

\begin{figure}[h]
  \centering
  \includegraphics[width=8cm]{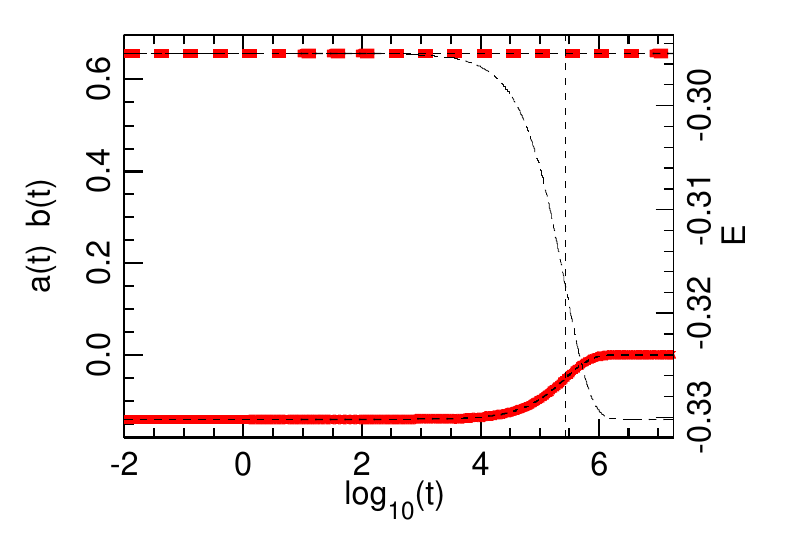}
  \caption{Amplitudes of the eigenmodes $a(t)$ and $b(t)$ during the aging
    stage at temperature $T_1$ in the protocol to observe the Kovacs effect.
  The thick red lines sow the results of the numerical integration with a full
line for $a(t)$ and a dash line for $b(t)$. The short-dash black lines
are the results of the analytical calculation. The vertical short-dash
line marks the
value of $\tau_1(T_1)$. The thin dash-dot line shows the variation of the
energy (right scale).}
  \label{fig:modeskaging}
\end{figure}

\medskip
Besides providing an analytical method to study the out-of-equilibrium
properties of a system, the decomposition of the dynamics on the normal modes
brings further understanding of the observations. This is clear for
the protocol used to study the
Kovacs effect. We start from an equilibrium state at $T^{\mathrm{ini}}=0.30$
and therefore at $t=0$ the probabilities are $P_i(t=0) =
P_i^{\mathrm{eq}}(T^{\mathrm{ini}})$. At this time the temperature is switched
to $T_1=0.02$. Solving Eq.~(\ref{eq:defvectpr}) we can determine the eigenvalues
$\lambda_{1,2}(T_1)$, i.e. the relaxation times $\tau_{1,2}(T_1) =
-1/\lambda_{1,2}(T_1)$ and the eigenvectors $\vec{U}^{(1)}(T_1)$,
$\vec{U}^{(2)}(T_1)$. This gives $\tau_{1}(T_1) = 0.2682\,10^6$t.u. and
$\tau_{2}(T_1) = 0.8708\,10^{12}$t.u.~.
Solving Eq.~(\ref{eq:decomposQ}) for the initial
condition $Q_i(t=0) = P_i(t=0) - P_i^{\mathrm{eq}}(T_1)$ we compute $a(t=0)$,
$b(t=0)$, and then Eq.~(\ref{eq:solab}) gives $a(t)$ and $b(t)$
at any time up to
the end of the aging period $t=t_1$. Figure \ref{fig:modeskaging} shows that
the analytical calculation perfectly agrees with the numerical integration, but
it also helps us understand what happens during the aging stage. Mode 1
i.e.\ $a(t)$, which turns out to be negative,
relaxes towards $0$ during aging, with the characteristic time $\tau_1$.
Note that, if the exponential relaxation of $a$ appears like a step, it is due
to logarithmic scale used for time on the horizontal axis. As
$t_1 = 1.8\,10^7 \ll \tau_2$ the amplitude $b(t)$ does not show any
noticeable change during aging. As shown by Fig.~\ref{fig:modeskaging} the
relaxation of $a(t)$ is associated to a significant energy drop. However
as the aging time was much shorter than $\tau_2$, the system has not reached
equilibrium at the end of aging.

\begin{figure}[h]
  \centering
  \includegraphics[width=8cm]{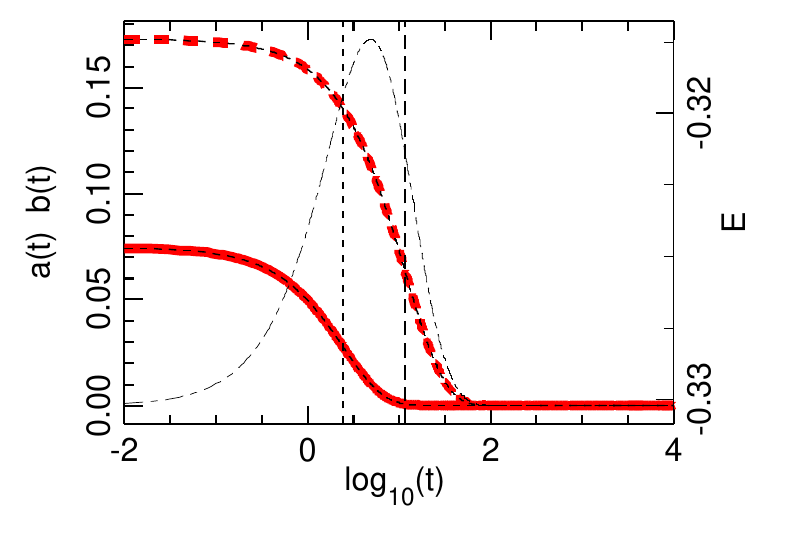}
  \caption{Amplitudes of the eigenmodes $a(t)$ and $b(t)$ during the
    constant temperature time evolution 
    at temperature $T_2$ in the protocol to observe the Kovacs effect.
    The thin dash-dot line shows the variation of the
    energy (right scale). It reproduces the result already shown in
    Fig;~\ref{fig:kovacs1}-a, which can now be understood in terms of
    normal modes.   
  The thick red lines sow the results of the numerical integration with a full
line for $a(t)$ and a long-dash line for $b(t)$. The short-dash black lines
are the results of the analytical calculation. The vertical lines marks the
values of $\tau_1(T_2)$ (short-dash line) and $\tau_2(T_2)$ (long-dash line). }
  \label{fig:modeskovacs}
\end{figure}

The analytical calculation gives the value of $a(t_1)$, $b(t_1)$ at the end of
the aging stage of the Kovacs protocol. Therefore we know the probabilities
$P_i(t_1)$, from which we can deduce the values $Q_i(t_1^+)$ of the deviations
with respect to $P_i^{\mathrm{eq}}(T_2)$ at the beginning of the last stage of
the Kovacs protocol, which is designated as $t_1^+$, to point out that we
consider this point as the start point of the next stage. Projecting the
$Q_i(t_1^+)$ on the normal modes {\em at temperature $T_2$} we get the
initial values $a_0(t_1^+)$,  $b_0(t_1^+)$ of the amplitudes of the eigenmodes
at the beginning of this last stage, and we obtain $a(t)$,
$b(t)$ for the full 
duration of the Kovacs stage, which gives us the probabilities and the energy
during this stage. This allows a full analytical calculation of the Kovacs
effect in the three-state model, but, as for the relaxation to equilibrium
discussed above, it also provides a
quantitative understanding of the properties of
the Kovacs hump. Figure \ref{fig:modeskovacs} shows that the
hump occurs because the 
two modes relax at different time scales. The relaxation times at
temperature $T_2$ are $\tau_1 = 2.471\;$t.u. and $\tau_2 = 11.59\;$t.u.~.
They differ by many orders of magnitude from the relaxation times
during aging. This is typical for a process taking place in the vicinity
of a glass transition, as in the experiments of Kovacs. The
rise of the Kovacs hump is observed when mode 1 relaxes, and the drop of the
hump is due to the relaxation of mode 2. Therefore the properties of the
Kovacs hump, shown in Fig.~\ref{fig:kovacs2} can be quantitatively explained
in terms of the eigenmodes of the three-state system. When $\Delta T =T_2 -
T_1$ decreases, i.e.\ $T_2$ decreases towards the low temperature $T_1$ for
which the relaxation times become very long, the Kovacs hump moves later in
time. Moreover, when the temperature decreases, the ratio $\tau_2/\tau_1$
increases. This explains why the hump becomes broader. The decay of its
amplitude is also easy to understand qualitatively. When $T_2$ gets closer to
$T_1$ the probabilities $P_i^{\mathrm{eq}}(T_2)$ are closer to
$P_i^{\mathrm{eq}}(T_1)$, which leads to smaller values of $Q_i(t_1^+)$. Then
the $a_0(t_1^+)$,  $b_0(t_1^+)$ are smaller,
so that the relaxation of the two modes,
which generates the Kovacs hump, is weaker.

\section{Extending thermodynamics to out-of-equilibrium systems ?}
\label{sec:fictiveT}

\subsection{Fictive temperature}

The concept of {\em fictive temperature} $T_f$, introduced by Tool and Eichling
\cite{TOOL1931} assumes that any state of an out-of-equilibrium system at
temperature $T$ is identical to an equilibrium state at another temperature
$T_f$. This is a powerful concept which has been widely used to study
glasses \cite{WONDRACZEK, GUO-2011, VALLE-2012},
however it has some limitations \cite{RITLAND,LEUZZI}.
This idea implies that an out-of-equilibrium system can be fully described by
a set of thermodynamic variables at equilibrium, and, as
discussed in the introduction, this 
is not compatible with the existence of the Kovacs effect. In the Kovacs
protocol, at the end of the aging stage, when energy has reached the value
$E_1$ we look for a temperature $T_2$ such that $E^{\mathrm{eq}}(T_2) =
E_1$. Using the concept of fictive temperature, it
means that $T_2 = T_f$ so that the state reached at
the end of aging should be identical to the equilibrium state at temperature
$T_2$. The experimental Kovacs hump, as well as its analysis within the
three-state system discussed above, show that it is not true.

\medskip
To understand the origin of this difficulty with the concept of $T_f$, it is
necessary to come back to the fundamental definition of temperature in
thermodynamics, expressed by the standard formula $dS = \delta Q / T$,
relating the entropy
$S$ and the heat exchange $\delta Q$ in an infinitesimal transformation
occurring at equilibrium.
For an out-of-equilibrium process, $1/T$ is no longer an integrating factor
which can turn the heat exchange $\delta Q$ into an exact differential form
because $\delta Q / T$ depends on the pathway in the configuration space, as
shown by the Clausius inequality for a cycle. Nevertheless one can define a
fictive temperature $T_f^S$ which plays the role of such an integrating factor
by extending the usual definition of temperature so that the integration of
$\delta Q/T_f^S$ along a temperature 
cycle vanishes  whatever the thermodynamic path in the configuration space,
i.e.\ whatever the thermal history of the system along the cycle.
For a system such as the three-state system
the heat exchange
is merely equal to the variation of its energy, so that
the fictive temperature should be defined by
  \begin{equation}
    \label{eq:defTS}
    \frac{1}{T_f^S} = \frac{\partial S}{\partial E} \; .
  \end{equation}
  Consequently, when the system undergoes an infinitesimal 
transformation in which the probabilities $P_i$ change by $dP_i$,
from the expression of the energy $E = \sum P_i \, E_i$ and the
statistical expression of the entropy $S = - \sum P_i \; \ln P_i$ (with $k_B=1$
because we express temperatures in energy units), we get
  \begin{equation}
    \label{eq:defTfs}
    T_f^S = - \; \frac{\sum_{i=1}^N E_i \; dP_i}{\sum_{i=1}^N \ln P_i
      \; dP_i} \; .
    \end{equation}
In the particular case of the three-state system with $P_3 = 1 - P_1 - P_2$,
this leads to
\begin{equation}
      \label{eq:defTfs3}
       T_f^S = - \; \frac{(E_1-E_3) dP_1 + (E_2 - E_3) dP_2}%
        {dP_1 \ln\left(\frac{P_1}{P_3}\right) +
          dP_2 \ln\left(\frac{P_2}{P_3}\right) }
\end{equation}

\medskip
For the three-state system, $P_1$ and $P_2$ are {\em independent} variables,
so that $dP_1$, $dP_2$ are arbitrary. Therefore Eq.~(\ref{eq:defTfs3}) shows
that the value of $T_f^S$ {\em cannot be defined in term of the microscopic
variables $P_1$, $P_2$. It also depends on the particular infinitesimal
transform which is considered,} i.e.\ on the path that the system follows in
its parameter space $P_1$, $P_2$. Let us label
$T_f^{S,\mathrm{transfo}}$ this value of
the fictive temperature to stress that is is only meaningful for a
particular transformation.

When the system approaches equilibrium, the occupation probabilities $P_i$
tend to $P_i = (1/Z) \exp(-E_i/T)$ and therefore the denominator of $T_f^S$ in
Eq.~(\ref{eq:defTfs3}) tends towards $\left[(E_3 - E_1) dP_1 + (E_3 - E_2)
  dP_2\right]/T$ so that the expression of $T_f^S$ simplifies and tends to $T$,
independently of $dP_1$, $dP_2$, i.e.\ whatever the path used to approach
equilibrium, as expected for a fictive temperature.

While $T_f^S$ cannot be uniquely defined from the
instantaneous state of a glassy system out of equilibrium, it is
nevertheless possible
to define fictive temperatures which have an intrinsic meaning.
{\em As we can
  define fictive temperatures associated to particular transformations, it is
interesting to consider the transformations associated to the eigenmodes.}
At fixed temperature we have $dP_i = dQ_i$ because the equilibrium
probabilities are constant. Moreover the time dependence of the $Q_i$s is
related to the eigenmodes by Eqs~(\ref{eq:decomposQ}) and (\ref{eq:solab})
i.e.\
\begin{align}
  \label{eq:qimode}
  Q_1 &= a_0 e^{-t/\tau_1} U_1^{(1)} + b_0 e^{-t/\tau_2} U_1^{(2)}
  \nonumber \\
  Q_2 &= a_0 e^{-t/\tau_1} U_2^{(1)} + b_0 e^{-t/\tau_2} U_2^{(2)} \; .
\end{align}
Therefore, if only mode 1 is excited ($b_0 = 0$), for a time step $dt$ we have
\begin{equation}
  \label{eq:dq1sdq2}
  \frac{dQ_1}{dQ_2} = \frac{dP_1}{dP_2} = \frac{U_1^{(1)}}{U_2^{(1)}} \; .
\end{equation}
(Remember that $dP_i = dQ_i$.)
The fictive temperature $T_f^S$ associated to this transformation,
that we denote $T_f^{S,1}$ because it corresponds to a transformation in which
mode 1 only is activated, is equal to
 \begin{equation}
      \label{eq:defTfs1}
      T_f^{S,1} = - \frac{(E_1-E_3) U_1^{(1)} + (E_2 - E_3) U_2^{(1)}}%
        {U_1^{(1)} \ln\left(\frac{P_1}{P_3}\right) +
          U_2^{(1)} \ln\left(\frac{P_2}{P_3}\right) } \; .
\end{equation}
A similar formula, involving the components $U_1^{(2)}$, $U_2^{(2)}$ of the
eigenvector of mode 2 defines $T_f^{S,2}$ for a transformation involving
mode 2 only, starting from the state $P_1$, $P_2$.

Therefore, for any time evolution of the system, for each state $P_1$, $P_2$ we
can define two fictive temperatures  $T_f^{S,1}$,  $T_f^{S,2}$ which tell us
how far the system is from its actual temperature $T$ {\em regarding one
  particular mode, i.e.\ regarding one particular relaxation time.}

\begin{figure}[h]
  \centering
  \begin{tabular}{c}
    (a) \\
    \includegraphics[width=7cm]{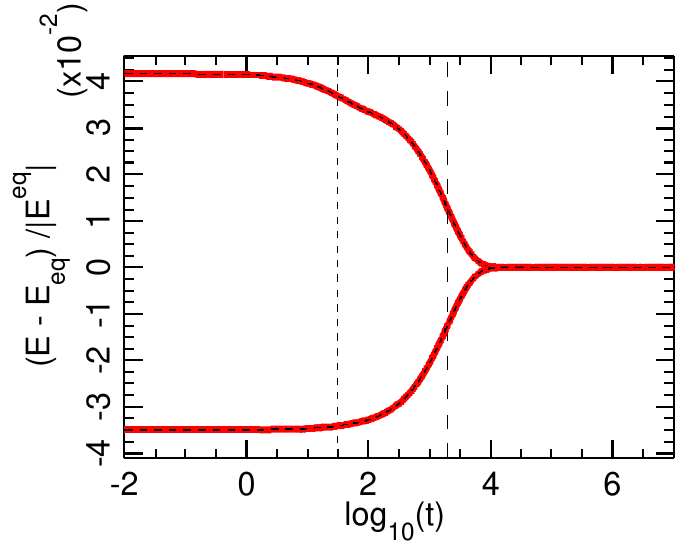} \\
    (b) \\
    \includegraphics[width=7cm]{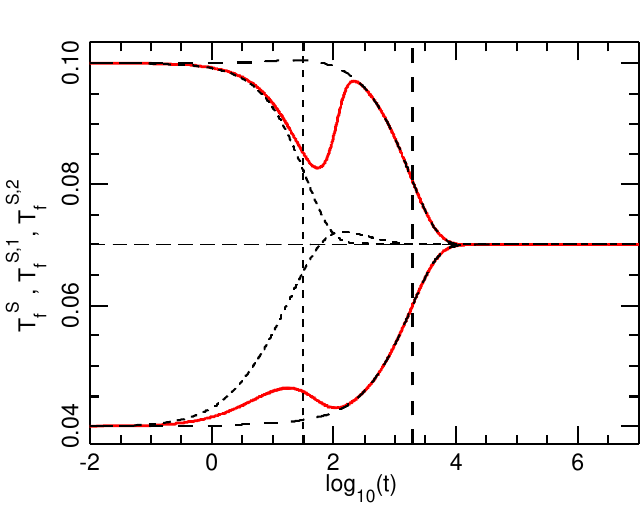}
  \end{tabular}
  \caption{Approach to equilibrium: the temperature of the three-state system
    in equilibrium at $T=0.10$ or $T=0.04$ was abruptly
    switched to $T=0.07$ and we
    followed the evolution of the system for $10^7\;$t.u.~.
    (a): Evolution of the energy of the system versus time. The figure shows
the deviation of the energy $E(t)$ from the equilibrium energy $E^{\mathrm{eq}}$
at $T=0.07$ divided by $E^{\mathrm{eq}}$.  The thick red lines are the
numerical results. The thin black lines superimposed to them are the results
of analytical calculations using the expansion on the eigenmodes.
    (b) Fictive temperatures during approach to equilibrium from above and
    from below. Red full line: $T_f^{S,\mathrm{transfo}}$
    computed along the thermodynamic trajectory.
    Short-dash  and long-dash black curves: $T_f^{S,1}$ and
    $T_f^{S,2}$, the fictive temperatures relative to the two
    eigenmodes. On both panels the vertical lines mark the
    relaxation times of the modes
    $\tau_1$ (short dash) and $\tau_2$ (long dash) at thermostat
    temperature $T$ marked by  the thin dashed
  horizontal line.} 
  \label{fig:approach-equil}
\end{figure}

As a first example let us study another characteristic feature of glasses,
which was exhibited by Kovacs in his pioneering work \cite{KOVACS}, the
asymmetry of the kinetics
of the approach to equilibrium upon cooling versus heating
in the range of the glass transition.  Figure \ref{fig:approach-equil} shows
the behavior of the three-state model when it approaches equilibrium at
$T=0.07$ both on cooling from an equilibrium state at $T=0.10$ and on heating
from $T=0.04$.  In each case, the temperature is abruptly changed from the
initial temperature to $T=0.07$. Figure \ref{fig:approach-equil}-a shows the
evolution of the deviation of the energy $E(t)$ from the equilibrium
energy $E^{\mathrm{eq}}$ at $T=0.07$. The variations on cooling and on heating
are not symmetric, with an energy change starting earlier on cooling than on
heating. As noticed by Kovacs, this effect is not trivial and it has been
reexamined recently in a study which analyzes it in terms of spatial
heterogeneities \cite{LULLI}.
Figure \ref{fig:approach-equil}-a shows that it can be observed in the
simpler three-state system and
Fig.~\ref{fig:approach-equil}-b shows how the evolution of the
fictive temperatures $T_f^{S,\mathrm{transfo}}$, $T_f^{S,1}$ and
$T_f^{S,2}$ can help to understand how the relaxation modes of the system
contribute. As the system starts and ends in equilibrium states,
the red curves of the fictive temperatures along the cooling and heating
processes $T_f^{S,\mathrm{transfo}}$ start and end at the values of the
initial and final thermodynamic temperatures. However their evolution is not
monotonic. Both show an intermediate extremum, which can be tracked to the
role of the two modes of the system. First $T_f^{S,\mathrm{transfo}}$ tends
to follow the fastest mode $T_f^{S,1}$ but the relaxation of this mode is not
sufficient to reach equilibrium and then $T_f^{S,\mathrm{transfo}}$ follows
the evolution of the slower mode $T_f^{S,2}$. However, as the initial states
at $T=0.04$ and $T=0.10$ were different, the initial amplitudes of the
modes $a(t=0)$,
$b(t=0)$ were not the same at the beginning of the cooling or
heating processes so that the relative
contribution of the two modes is not the same in the two processes. This
gives rise to an effective relaxation rate which is different in the two
cases. It is interesting that this peculiarity can be observed in a simple
model without disorder. Instead of spatial inhomogeneities, the different
trajectories in the configuration space lead to different values of
$T_f^{S,\mathrm{transfo}}$ which play a similar role.

\begin{figure}[h]
  \centering
  \begin{tabular}{c}
    (a) \\
    \includegraphics[width=7cm]{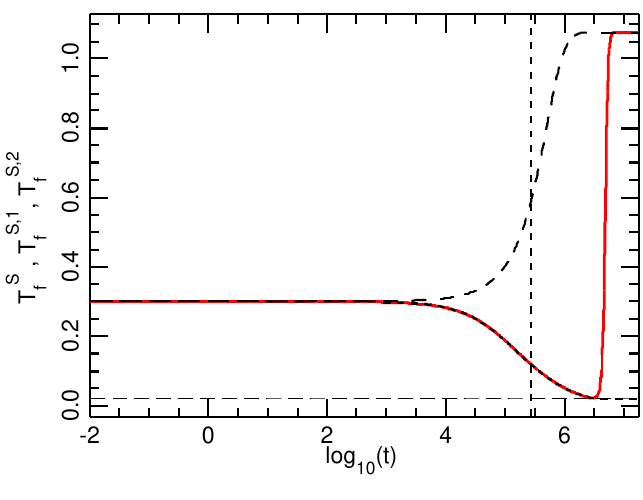} \\
    (b) \\
    \includegraphics[width=7cm]{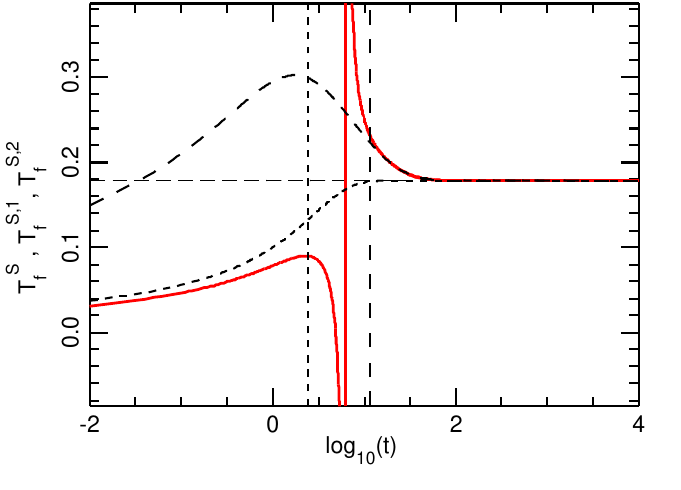}
  \end{tabular}
  \caption{Fictive temperatures during the two stages of the Kovacs protocol:
    (a) aging at $T_1$, (b) time evolution at $T_2$ leading to the Kovacs
    hump. On panel (b) the vertical scale has been truncated to allow a better
    view of the behavior of the fictive temperatures relative to the
    eigenmodes. 
    Red full line: $T_f^{S,\mathrm{transfo}}$
    computed along the thermodynamic trajectory.
    Short-dash  and long-dash black curves: $T_f^{S,1}$ and
    $T_f^{S,2}$, the fictive temperatures relative to the two
    eigenmodes. The vertical lines mark the relaxation times of the modes
    $\tau_1$ (short dash) and $\tau_2$ (long dash) at each thermostat
    temperature $T$ marked by  the thin dashed
  horizontal line.} 
  \label{fig:tfict-kovacs}
\end{figure}

Figure \ref{fig:tfict-kovacs} shows that the concept of fictive
temperature is also useful to
understand the time evolution of the system during the two stages of the
Kovacs protocol.
For the aging stage (Fig.~\ref{fig:tfict-kovacs}-a) we notice that all fictive
temperatures $T_f^{S,\mathrm{transfo}}$, $T_f^{S,1}$, $T_f^{S,2}$, start from
the value $0.3$ which was the temperature of the equilibrium state before the
system was quenched to $T_1$. They start evolving around $t= \tau_1(T_1)$,
which is the time necessary for the relaxation of the fastest mode to show
up. In this time range the $P_i$ start to evolve, and, as a result, all
fictive temperatures change. $T_f^{S,\mathrm{transfo}}$ closely follows the
fictive temperature $T_f^{S,1}$ because, at this low temperature $T_2=0.02$,
the two relaxation times are extremely different. As shown in
Fig.~\ref{fig:modeskaging}, only mode 1 significantly evolves during the aging
process. Therefore the path in the configuration space follows mode 1, so
that  $T_f^{S,\mathrm{transfo}} \approx T_f^{S,1}$. After mode 1 has
equilibrated, its amplitude $a(t)$ has almost vanished and the evolution goes on
along mode 2 only. Even though the variation of $b(t)$ is small, it is now the
dominant mode in the system and therefore $T_f^{S,\mathrm{transfo}} \approx
T_f^{S,2}$. The switch of $T_f^{S,\mathrm{transfo}}$ from $T_f^{S,1}$ to
$T_f^{S,2}$ looks very sharp on Fig.~\ref{fig:tfict-kovacs}-a but this is an
artifact of the logarithmic scale used for time. Therefore the plot of the
fictive temperatures allows a detailed analysis of the mechanisms behind the
aging process. At the end of this stage, $T_f^{S,\mathrm{transfo}}$ is very
different from the temperature of the thermostat, which tells us that the
system is still strongly out of equilibrium
in spite of this long aging
period. Fig.~\ref{fig:tfict-kovacs}-b allows a similar analysis for the process
leading to the Kovacs hump, but, in this case the evolution is more complex.
When the temperature is switched from $T_1$ to $T_2$, the eigenvectors of the
modes change, and therefore the expressions of $T_f^{S,1}$, $T_f^{S,2}$ change
too. The abrupt temperature rise from $T=0.02$ to
$T=0.17856$ induces a change in the occupation of the three states
(see Fig.~\ref{fig:kovacs1}-b). Initially $P_3$, was almost zero, giving
very large  ratios $P_1/P_3$ and $P_2/P_3$. As it
raises significantly, the ratios decrease as well as 
the logarithms  in  the denominators of   $T_f^{S,1}$, $T_f^{S,2}$. As a result
both fictive temperatures tend to grow for the shorter times.
$T_f^{S,1}$ smoothly approaches the system temperature $T_2$
while the initial growth of $T_f^{S,2}$ is large enough to bring it above
$T_2$ and it finally reaches $T_2$ by decaying from above. For the Kovacs
stage, the evolution of $T_f^{S,\mathrm{transfo}}$ is very peculiar, and this
points out the complexity of the Kovacs relaxation. Initially, as mode 1 is
the fastest, it dominates the evolution and therefore
$T_f^{S,\mathrm{transfo}}$ evolves like  $T_f^{S,1}$. In the last part of the
process, as mode 1 has relaxed, $T_f^{S,\mathrm{transfo}}$ evolves like
$T_f^{S,2}$ as in the aging period. But, in between, around the maximum of
the Kovacs hump between $\tau_1(T_2)$ and $\tau_2(T_2)$,
$T_f^{S,\mathrm{transfo}}$ shows an unexpected behavior, with a divergence and
even a short time interval where it is {\em negative}. To understand it, one
has to look at the definition of the fictive temperature given by
Eq.~(\ref{eq:defTS}) and at the behavior of the energy
(Fig.~\ref{fig:kovacs1}) and of the entropy discussed below
(Fig.~\ref{fig:entropy-kovacs}). Both show a hump. At the maximum of the
Kovacs hump we have $dE = 0$, and it changes sign when the Kovacs hump
occurs. Similarly the entropy shows a hump with $dS = 0$, and a change of sign
of $dS$ in the vicinity. But, due to the entropy production
(Sec.~\ref{sec:entropy}) the maximum of the entropy does not exactly coincide
with the maximum of the energy. The vanishing of $dS$ when $dE \not= 0$ causes
the divergence seen on $T_f^{S,\mathrm{transfo}}$, and the changes of signs of
$dE$ and $dS$ explain the surprising negative fictive temperature
$T_f^{S,\mathrm{transfo}}$.

\medskip
This analysis confirms that a unique
fictive temperature cannot be defined for an out-of-equilibrium system
which requires several configuration parameters besides
its thermodynamic variables to be fully characterized.  Nevertheless this
concept can be defined for a specific transformation, and particularly for
the eigenmodes. This gives a measurement of the distance to equilibrium
regarding each relaxation time of the system. Following
$T_f^{S,\mathrm{transfo}}$ and its relationship with the fictive temperatures
associated to each mode
gives a further understanding of the behavior of an out-of-equilibrium system.

\bigskip
\noindent{\em Particular case of the two-state system}

The same thermodynamic definition of the fictive temperature can also be
applied to the two-state system of Fig.~\ref{fig:model}-b. Using
$P_1 + P_2 = 1$, and therefore $dP_2 = - dP_1$, Eq.~\ref{eq:defTfs} reduces to
\begin{equation}
  \label{eq:Tfs2st2}
  T_f^S = - \frac{E_1 - E_2}{\ln \left( \frac{P_1}{P_2} \right) } \; .
\end{equation}
For this system the fictive temperature {\em does not depend on the particular
  transformation which is considered} because $dP_1$ does not appear in its
expression. This should not be a surprise because the state of this simple
system is fully described by one variable, which could be either $P_1$ or the
thermodynamic temperature, whether the system is in equilibrium or not. As
we noticed, this two-state system does not show the Kovacs effect. In this case
the temperature $T_2$ for which $E_1 = E^{\mathrm{eq}}(T_2)$ is the (unique)
fictive temperature. It means that the state reached after the aging state,
when the system is brought to $T_2$, is already its equilibrium state so that
it does not evolve any more.

\subsection{Entropy}
\label{sec:entropy}

The statistical
entropy of a model based on a set of energy states $E_i$, occupied with
probabilities $P_i$, over which the system evolves by transitions from a state
to another is given by
\begin{align}
  \label{eq:defS}
  {S} = - \sum_i P_i \; \ln P_i
\end{align}
where we have set $k_B=1$ because we measure temperatures in energy units. For
the three-state system it can be computed exactly in a numerical simulation
and even calculated analytically for all transformations involving only
constant-temperature processes and sharp temperature jumps. However, for a
discussion of glassy properties, it would be useful to get an expression of
the entropy which provides a deeper insight by distinguishing the equilibrium
contribution from the part coming from out-of-equilibrium
transformations. This can be done by exhibiting the equilibrium values within
$P_i$ using
\begin{equation}
  \label{eq:piq}
  P_i(T,t) = P_i^{\mathrm{eq}}(T) + Q_i(T,t) \; ,
\end{equation}
as we did earlier. For any transformation which is not too far from
equilibrium, i.e.\ such that $Q_i/P_i^{\mathrm{eq}} \ll 1$ then $\ln P_i$ can
be expanded as
\begin{align}
  \label{eq:devlog}
  \ln P_i &= \ln P_i^{\mathrm{eq}} + \ln \left(1 +
            \frac{Q_i}{P_i^{\mathrm{eq}}}\right) \nonumber \\
          &= \ln P_i^{\mathrm{eq}} + \frac{Q_i}{P_i^{\mathrm{eq}}} - \frac{1}{2}
            \left[ \frac{Q_i}{P_i^{\mathrm{eq}}} \right]^2 +
            O[(Q_i/P_i^{\mathrm{eq}})^3] \; .
\end{align}
Using Eq.(\ref{eq:defS}), this leads to an approximate expression of the
entropy as  
\begin{align}
  \label{eq:devS1}
  {S}= &-\sum_i P_i^{\mathrm{eq}} \; \ln P_i^{\mathrm{eq}}
  \nonumber \\
                    &- \sum_i Q_i \ln P_i^{\mathrm{eq}} 
                    - \frac{1}{2} \sum_i P_i^{\mathrm{eq}}\left[\frac
                      {Q_i}{P_i^{\mathrm{eq}}}\right]^2 
                   \nonumber \\   &+ O[(Q_i/P_i^{\mathrm{eq}})^3] \; .
\end{align}
which can be written
\begin{align}
  \label{eq:devS2}
  {S} = {S^{\mathrm{eq}}}+\Delta S_{1}+\Delta S_{2}+
  O[(Q_i/P_i^{\mathrm{eq}})^3] \; . 
\end{align}
where $S^{\mathrm{eq}}$ is the entropy at equilibrium at temperature
$T$. Taking into account  $\sum_i Q_i = 0$ and introducing the values
(\ref{eq:pieq}) of $P_i^{\mathrm{eq}}$, the correction to the entropy up to
first order in $Q_i/P_i^{\mathrm{eq}}$ is simply 
\begin{align}
  \label{eq:Sfirstorder}
  {\Delta S_{1}} = - \sum_i Q_i \ln P_i^{\mathrm{eq}} = \frac{ E -
  E^{\mathrm{eq}}}{T} \; . 
\end{align}
The second order correction to the entropy is
\begin{align}
  \label{eq:Ssecondorder}
  {\Delta S_{2}} = -\frac{1}{2}\left\langle \left[
  \frac{Q_i}{P_i^{\mathrm{eq}}} \right]^2\right\rangle \; . 
\end{align}
where the bracket means the statistical averaging.
 
For the three-state system, the different expressions
of the entropy can be computed analytically for any transformation
at constant temperature, starting from a known state $(P_1,P_2)$,
as shown in Sec.~\ref{sec:analytical}.

\begin{figure}[h]
  \centering
  \includegraphics[width=8cm]{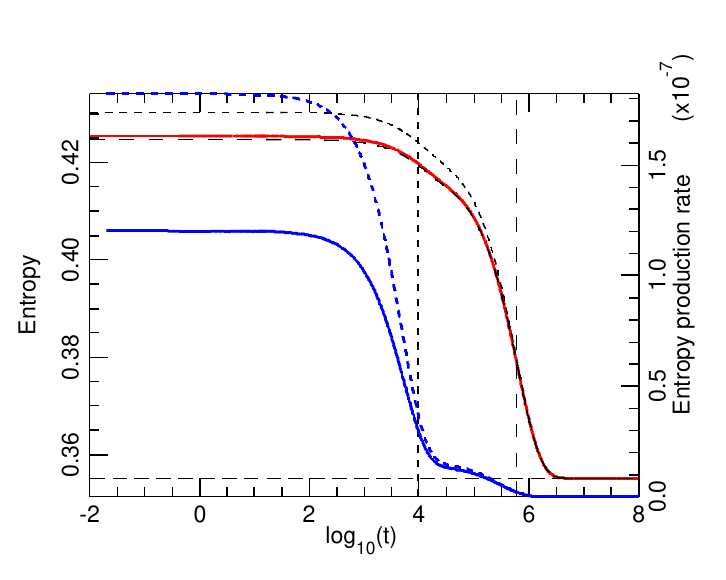}
  \caption{Left scale: entropy for the approach to equilibrium in an abrupt
    temperature change from $T=0.08$ to $T=0.07$.
    Red curve: Numerical result from Eq.~(\ref{eq:defS})
    (which could also be obtained
    analytically as we can compute $P_i(t)$).
    Short-dash and long-dash black lines: first and second order expansions.
    Note
that, in this case which does not include a big $T$-jump the second order
expansion for the entropy is accurate.\\
Right scale: Entropy production rate for the same process. Blue full line:
exact result from Eq.~(\ref{eq:entropycreation}) and dashed blue line:
    approximate value from Eq.~(\ref{eq:entropy-production}). \\
The vertical lines mark the relaxation times
    (full line $\tau_1$, long-dash line $\tau_2$).
} 
  \label{fig:entropy-approach}
\end{figure}

Figure \ref{fig:entropy-approach} shows the variation of the entropy when the
three-state system approaches equilibrium at $T=0.07$ from an initial
equilibrium at $T=0.08$, chosen sufficiently close to the
final temperature to avoid large deviations of the probabilities from their
equilibrium values to guarantee the validity of the expansion. The red
curve shows the exact value of the entropy. The first order approximation
\begin{align}
  \label{eq:devS12}
  {S} = {S^{\mathrm{eq}}}+\Delta S_{1} \; .
\end{align}
(small-dash black curve) overestimates the entropy, but the
second order expansion 
\begin{align}
  \label{eq:devS22}
  {S} = {S^{\mathrm{eq}}}+\Delta S_{1}+\Delta S_{2} \; .
\end{align}
(long-dash black curve)
is able to closely approach the actual entropy change in the process.

\begin{figure}[h]
  \centering
  \includegraphics[width=8cm]{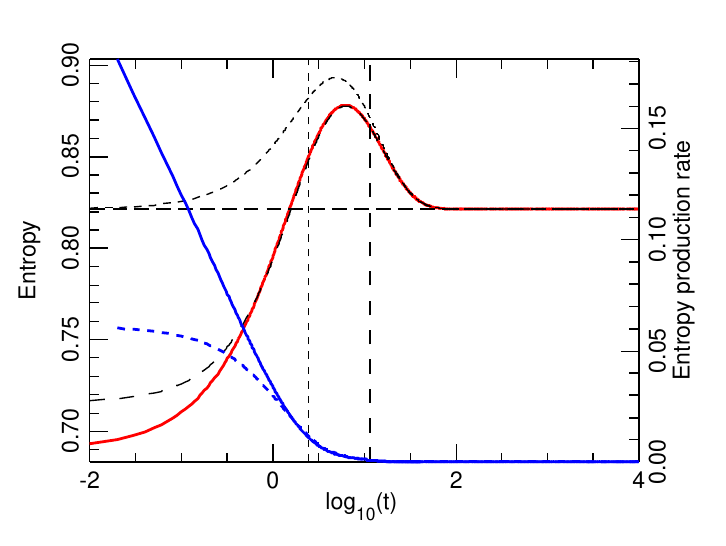}
  \caption{Left scale: entropy for the Kovacs time evolution.
    Red curve: Numerical result from Eq.~(\ref{eq:defS})
    (which could also be obtained
    analytically as we can compute $P_i(t)$).
    Short-dash and long-dash black lines: first and second order expansions.
In this case the calculation is less accurate than for the case shown in
  Fig.~\ref{fig:entropy-approach} because the condition $Q_i/P_i^{\mathrm{eq}}
  \ll 1$ is not satisfied in the first part of the
  time evolution. \\
Right scale: Entropy production rate for the same process. Blue full line:
exact result from Eq.~(\ref{eq:entropycreation}) and dashed blue line:
    approximate value from Eq.~(\ref{eq:entropy-production}). \\
The vertical lines mark the relaxation times
    (full line $\tau_1$, long-dash line $\tau_2$).
  } 
  \label{fig:entropy-kovacs}
\end{figure}

Figure \ref{fig:entropy-kovacs} shows the
same entropy expansion for last stage of the Kovacs protocol, at constant
temperature $T_2$, in which the 
energy exhibits the Kovacs hump plotted in Fig.~\ref{fig:kovacs1}.
The exact
calculation (red curve) shows that the entropy increases during the
process. The first order expansion of Eq.~(\ref{eq:devS1}) shows a hump in
the entropy but no overall entropy change in the whole process,
in agreement with Eq.~(\ref{eq:Sfirstorder}), because at the
end of the process the energy recovers the value that it had the beginning.
The second
order correction detects the entropy increase, however it does
not give an accurate evaluation of the entropy in the early stage of the
evolution. This is due to the large temperature jump, from $T_=0.02$ to
$T_2 = 0.17856$ after which the condition  $Q_i/P_i^{\mathrm{eq}} \ll 1$ is
not satisfied.

\subsubsection*{Entropy production}

The variation of the entropy in a transformation can be separated into two
terms
\begin{equation}
  \label{eq:dSei}
  \Delta S = \Delta_e S + \Delta_i S \; ,
\end{equation}
where $\Delta_e S = \Delta E / T$ is the contribution coming from the energy
exchange with the outside and $\Delta_i S$ is the contribution coming from
internal transformations in the system, called the entropy production, which
is never negative $\Delta_i S \ge 0$ and vanishes for reversible
processes. This concept, familiar for the thermodynamic entropy in
out-of-equilibrium system \cite{GLANSDORFF},
also extends to the statistical entropy in
multilevel systems \cite{LANGER}.
For a transformation at constant temperature, 
$\Delta_i S $ is related to the variation of the thermodynamic
function free energy $F = E - T S$ because $\Delta F / T = \Delta E / T -
\Delta S$ so that $\Delta_i S = - \Delta F / T$.

The approximate expression of the entropy, derived above, can also be used to
express the rate of entropy production $d_iS/dt$ for a transformation starting
from an equilibrium state because Eq.~(\ref{eq:Sfirstorder}) shows that
$\Delta S_1 =  (E -  E^{\mathrm{eq}})/T = \Delta_e S$ and therefore
$ \Delta_i S = \Delta S_2 + O[(Q_i/P_i^{\mathrm{eq}})^3]$. 
Taking the time derivative of this equation, and using
Eq.~(\ref{eq:Ssecondorder}) gives 
\begin{align}
  \label{eq:entropy-production}
  {\frac{d_{i}S}{dt}} = -\frac{1}{T}\frac{dF}{dt} \approx -\sum_i \frac
  {Q_i}{P_i^{\mathrm{eq}}}\frac {dQ_i}{dt} > 0  \; , 
\end{align}
up to second order in $Q_i/P_i^{\mathrm{eq}}$. This value, which can be
computed analytically as shown in Sec.~\ref{sec:analytical}, can be compared
with the exact value of ${d_{i}S}/{dt}$ deduced from the expression
(\ref{eq:defS}) of the entropy by subtracting
$d_eS = - \sum \ln P_i^{\mathrm{eq}} \; dP_i$ which gives
\begin{align}
  \label{eq:entropycreation}
  \frac{d_{i}S}{dt} = -\sum_i
  \ln\left(\frac{P_{i}}{P_i^{\mathrm{eq}}}\right)\frac{dP_{i}}{dt} 
\end{align}

For the three-state system, figures \ref{fig:entropy-approach} and
\ref{fig:entropy-kovacs} show
the exact and approximated expressions of the entropy creation rate
as a function of time for the approach to equilibrium
 (Fig.~\ref{fig:entropy-approach}) 
 and the constant temperature evolution at temperature
$T_2$ in the protocol to observe the Kovacs effect
(Fig.~\ref{fig:entropy-kovacs}).
As expected, the entropy production rate is positive at the beginning of
relaxation when the system is outside equilibrium. It then tends towards zero
with different shapes for the two different relaxational processes until the
system reaches equilibrium for longer time. The full-line blue curves
(exact values from
Eq.~(\ref{eq:entropycreation})~) or the dashed-lines blue curves
(approximate expressions from
Eq.~(\ref{eq:entropy-production})~) differ significantly at the earlier stages
of each of the transformations. This is because, after the temperature jumps
that started the processes, the second order expansion in
$Q_i/P_i^{\mathrm{eq}}$ is not sufficient. This is obvious in the case
of Fig.~\ref{fig:entropy-kovacs} because the plots of the entropy
(red full line and black dashed lines on the same figure)
show that even its second order expansion of the entropy
is not good in the early
stage of the transformation. This is less obvious for
Fig.~\ref{fig:entropy-approach} because the plots of the entropy
(red full line and black dashed lines) show
that the approximation of the entropy up to the second order in
$Q_i/P_i^{\mathrm{eq}}$ is not bad. However the entropy production is small,
and results from a difference between two nearly equal terms
$\Delta S$ and $\Delta_e S$, so that even a small error in the
calculation of  $\Delta_e S$ can have a very significant effect.
Since the entropy production term is
always a term of second order in the departure from equilibrium, it is
generally small. Accurately fulfilling the criterion
$Q_i/P_i^{\mathrm{eq}} \ll 1$ is of great importance for such a small
contribution. It is less critical for contributions
of first order such as the entropy. In
calculations, as well as in experiments \cite{JABRAOUI}, determining the
entropy production is challenging. 

\smallskip
For the  protocol to observe the Kovacs effect,
as expected the Kovacs hump in energy shows up in the entropy change through
$\Delta_e S$, but the red curve for the entropy computed by
${S} = - \sum P_i \; \ln P_i$
shows a large additional contribution which results in a
significant rise of the entropy. This is due to the irreversibility of the
process. The blue curve shows that the (positive) entropy production rate is
particularly strong at the beginning of the time evolution, which is the
source of the large entropy rise. Then the system evolves towards equilibrium
and the entropy production rate drops. When time has reached $\tau_2(T_2)$, the
entropy production has almost disappeared, as expected in equilibrium, and the
entropy change is almost determined by the contribution coming from the energy
exchange.

\section{Discussion}
\label{sec:discussion}

The Kovacs effect is a nice example of out-of-equilibrium process because it
clearly shows that the state of such a system cannot be described only by its
thermodynamic equilibrium variables.
This effect is usually considered as typical of glasses which have a
continuum of relaxation times
but it is actually more generic. It can be observed in a
system as simple as the three-state system, with only two relaxation
times. However it does not exist for the even simpler two-state system. This
two-state system can show some glass-like properties, such as a negative
heat capacity \cite{BISQUERT},
however it lacks a fundamental property because, when its energy is specified,
its configuration, described by the unique variable $P_1$ is also fully
determined. In the three-state system this one-to-one correspondence between
the energy and the configuration of the system is broken. A given energy can
be realized by various $P_1$, $P_2$ configurations.

\medskip
The three-state system is an interesting model to study out-of-equilibrium
properties because it is probably the simplest system which has the necessary
ingredients. It is too simple to describe the glass transition by itself, but
it nevertheless exhibits fundamental properties of glasses, such as the Kovacs
effect. And moreover it allows an analytical description which leads to a
basic understanding of the phenomena. For instance the time at which the
Kovacs hump is observed can be clearly linked to the relaxation times of the
system. This provides some answer to a question raised by Bertin et al. in
ref.~\cite{BERTIN}: how to extract microscopic information on a system from
the Kovacs hump? Our results show that its shape
contains direct information on the
statistics of the relaxation times in the system.

\begin{figure}[h]
  \centering
  \includegraphics[width=8cm]{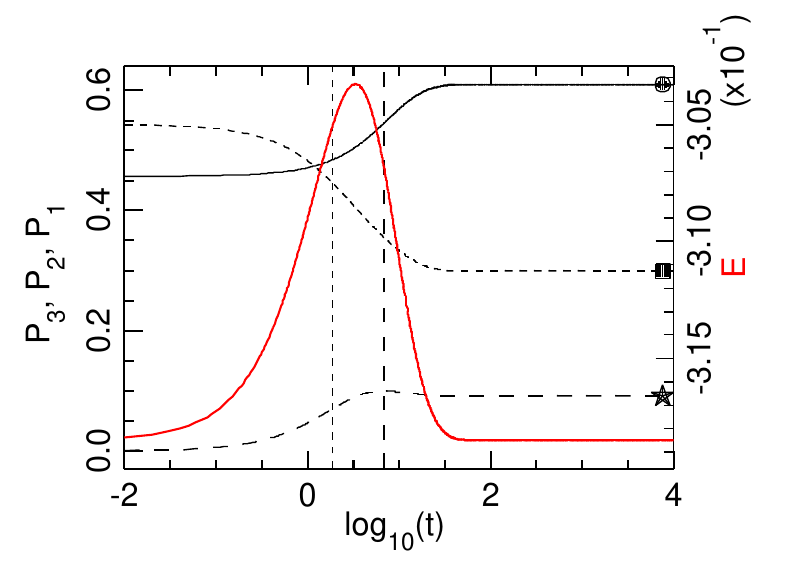}
  \caption{Energy (red curve, right scale) and probabilities $P_1$,$P_2$, $P_3$
    (black curves, left scale) versus
    time (in log scale) during the time evolution
    at constant temperature $T_2 = 0.2110$ after aging for $1.8\,10^7\,$t.u.
    at temperature $T_1 = 0.02$ in the study of the Kovacs effects.
    The conditions are very similar to the case shown in Fig.~\ref{fig:kovacs1}
    except that the aging started from an equilibrium temperature
    $T^{\mathrm{ini}} = 0.50$ instead of $0.30$ so that the cooling rate was
    faster. Full black line $P_1$, short-dash black line $P_2$,
    long-dash black line $P_3$.
    The points at the right side of the graph show the values of the
    equilibrium probabilities at $T_2$ : circle $P_1^{\mathrm{eq}}$,
    square $P_2^{\mathrm{eq}}$, star $P_3^{\mathrm{eq}}$.}
  \label{fig:far-equil}
\end{figure}

\medskip
One may wonder whether the three-state model can be relevant
  to describe some properties of complex non-equilibrium systems. An
  interesting example is provided by spin models of strong and fragile glasses
  studied in Ref.~\cite{ARENZON}.
This paper points out that the Kovacs effect seems to provide an alternative
independent method to obtain the equilibration time of the system because the
time at which the peak occurs varies like this equilibration time. This is
exactly what the three-state model gives by showing that the Kovacs hump rises
after the shortest relaxation time of the system and drops after the longest
relaxation time. The paper~\cite{ARENZON} makes a closer
connection to the ideas behind the three-state model because it studies how a
complex spin systems explores its inherent structures, i.e. the basins of
attraction of the energy minima. It shows that the distribution of the inherent
structures explored far from equilibrium differs in depth from the distribution
explored in equilibrium. A similar analysis was made in a study of the Kovacs
effect in a molecular liquid \cite{MOSSA}. It found that, when the system ages
it explores regions of the potential energy landscape {\em which are no
  explored at equilibrium}. The three-state model shows the same phenomena
when a fast quenching brings it very far from equilibrium, as shown in
Fig.~\ref{fig:far-equil} for an example of the Kovacs effect in which the
aging started from $T=0.50$ instead of $T=0.30$ as in the cases discussed
earlier. The largest temperature jump brought the system farther from
equilibrium when the temperature evolution at constant temperature
($T=0.2110$) started. The plot of the energy versus time on
Fig.~\ref{fig:far-equil} shows the usual hump, but the variation of the
occupation probabilities $P_1$, $P_2$, $P_3$ versus time shows a peculiar
evolution because two curves cross each other. At the end of aging $P_1$ (for
$E_1 = -0.40$) was {\em lower} than $P_2$ (for $E_2 = -0.25$). In equilibrium
we don't expect that an energy level is more
populated than another one which has a lower energy, but this can be observed
if the system is far from equilibrium. As observed in complex systems, the
three-state model explores such states.

It seems that a model as simple as the three-state model 
cannot quantitatively describe the properties of actual glassy samples but it
might actually be 
closer to experimental systems than one could think. In many glassy systems,
such as polymers, instead of a continuum spectrum of relaxation times, there
are groups of ``slow modes'' for instance associated to molecular
reorientations, which are well separated from ``fast modes''.
In Ref~\cite{PEREZ-EULATE},
long aging experiments on polymers at temperatures lower than the glass
transition temperature
show that, at such low temperatures,
it is possible to reach a plateau in enthalpy (by a fast 
relaxation mechanism) without disturbing the classical alpha relaxation
processes. It results in an aging-time-dependent
overshoot of the
specific heat during heating, below the classical specific heat peak at
the glass-transition temperature.
The three-level system, with a relaxation of the fast mode at low temperature
while the slow one is practically not disturbed
(see Fig.~\ref{fig:modeskaging}), could be
appropriate to explain this behavior. In this case two fictive
temperatures such as those associated to each independent modes of the three
level system should be involved. There are several experimental results which
point to the interest of this simple model to analyze Kovacs-hump data on
glasses. A recent example is provided by an article showing that
the Kovacs hump is not only observed in the enthalpy of a glass but can also
be detected in the amplitude of the Boson peak which corresponds to an enhanced
low frequency (terahertz region) density of states as compared with the
Debye square frequency law \cite{LUO-P}. This may be related to our
Fig.~\ref{fig:tfict-kovacs}-b which shows a peak in the fictive temperature of
the slow mode $T_f^{S,2}$, occurring near the time at which the Kovacs hump
occurs, while $T_f^{S,1}$ for the fast mode increases monotonically.
The paper suggests that a possible explanation of the experimental results
could rely on a two-relaxation-time model already used earlier to propose a
phenomenological analysis of Kovacs-hump-like non-monotonic relaxations of
the Curie temperature in a metallic glass \cite{GREER}
and refractive index of $B_2O_3$. In these studies the frequencies of the two
modes were simply fitted. It is tempting to try to go further with
the three-state model which allows the computation of the shape of the Kovacs
hump in addition to the determination of the relaxation times.

\medskip
\begin{figure}[h]
  \centering
  \includegraphics[width=8cm]{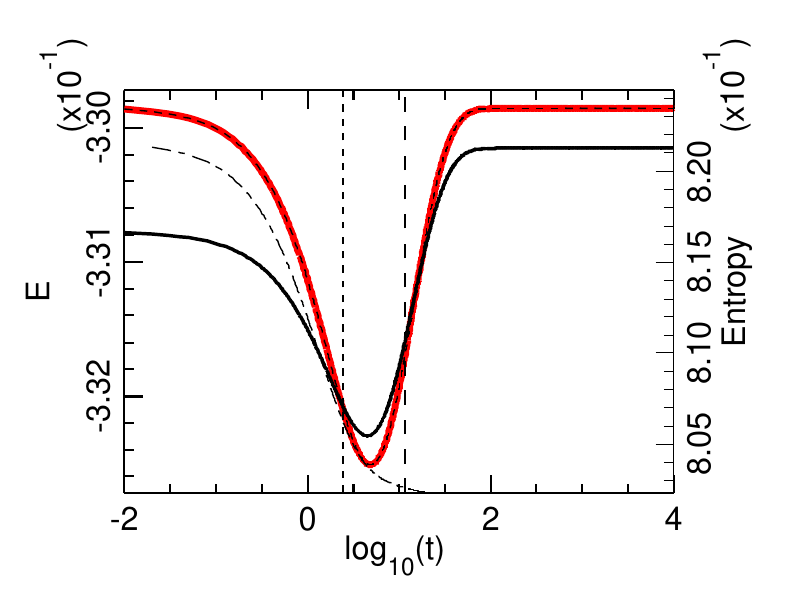}
  \caption{Reverse-Kovacs effect observed with the three-state model with
    $T'_0=0.02$, $T'_1 = 0.22$ and $T'_2=0.17852$. The red curve (numerical
    result) superimposed on the short-dash black line (theoretical
    calculation) shows the variation of the energy (left scale)
    versus time (in log scale) during the time evolution
    at temperature $T'_2$ after annealing for 10.12 t.u. at temperature $T'_1$.
    The full black line shows the time evolution of the entropy (right scale)
    and the thin dash-dot line shows the variation of the entropy production
    (its scale, not shown on the graph, varies from 0 to $3.0\,10^{-3})$}
  \label{fig:reversekovacs}
\end{figure}
In their experimental study of $B_2O_3$, Boesch et al.\
  exhibited another memory effect which is a kind of ``reverse-Kovacs
  effect'', later observed in other non-equilibrium systems
  \cite{TRIZAC,KURSTEN}. The 
  protocol is to start from an equilibrium state at {\em low temperature}
  $T'_0$, and then to anneal the sample at a higher temperature $T'_1$ for a
  time so short that it cannot reach equilibrium. The annealing is
  interrupted when the refractive index reaches a prescribed value which is
  its value at equilibrium at a temperature $T'_2 < T'_1$. Then the temperature
  is abruptly shifted to $T'_2$ and the behavior of the refractive index is
  measured. As for the Kovacs effect, although the index already has its
  expected value in equilibrium at $T'_2$, it does not stay constant but
  exhibits a non-monotonic evolution with the {\em opposite sign of its
  evolution in the Kovacs effect,} until it finally recovers its equilibrium
  value at temperature $T'_2$. As shown in Fig.~\ref{fig:reversekovacs}, this
  peculiar memory effect can also be observed with the three-state model with
  a similar protocol. We selected $T'_0=0.02$, $T'_1=0.22$ and an annealing time
  of $10.12\;$t.u.\ at $T'_1$ so that the energy reaches the equilibrium value
  $E^{\mathrm{eq}}(T'_2=0.17852)$, a temperature very close to the temperature
  $T_2=0.17856$ at which we had 
  studied the Kovacs effect plotted on Fig.~\ref{fig:kovacs1}. The time
  evolution of the energy at $T'_2$ shows {\em a dip} instead of the Kovacs
  hump. As we have chosen a value $T'_2$ very close to the temperature
  $T_2$ at which
  we studied the Kovacs effect shown in Fig.~\ref{fig:kovacs1}, the relaxation
  times are almost the same in the two cases and the shape of the dip looks
  as a mirror image of the Kovacs hump. However its magnitude
  is smaller than the magnitude of the hump because the annealing at $T'_1=0.22$
  brought the system closer to equilibrium than the annealing at $T_1=0.02$ 
  used to get the Kovacs effect. In this reverse-Kovacs effect, because the
  energy drops in the dip, the entropy shows a minimum (see
  Fig.~\ref{fig:reversekovacs}), but, as expected for this irreversible
  process the entropy production is always positive and the entropy during
  the evolution at $T'_2$ increases. Together with the asymmetric approach to
  equilibrium discussed earlier, this example shows the ability of the
  three-state model to describe a large variety of phenomena experimentally
  observed in glasses.

\medskip
It is interesting to examine whether the three-state system
contains the ingredients required
by Bouchbinder and Langer to model the Kovacs effect
\cite{BOUCHBINDER}. Their basic assumption is that a glass consists of two
interacting subsystems which have very
different characteristic times and can fall out of
thermal equilibrium with each other. 
The two eigenmodes of the three-state system play this role.
Figure \ref{fig:modeskaging}, plotting the amplitude of these modes at the end
of the aging period in the Kovacs protocol shows that one mode has already
relaxed to equilibrium while the second has not. Thus the two modes can
indeed fall out of thermal equilibrium with each other. We have shown that we
can affect a fictive temperature to each mode, and there are states
in which the two fictive temperatures are different. The third, and less
obvious condition that there exists a mechanism in which the fast subsystem
can produce changes in the effective temperature of the slow subsystem seems
to be absent in the three-state model because the eigenmodes are independent
from each other. However because the dynamics of the model is driven by the
thermal bath which is responsible for the evolution of the state populations,
this is enough to ensure that the slow subsystem keeps evolving even when the
fast mode is equilibrated, although it cannot be attributed to a weak coupling
between the two eigenmodes.
In spite of the similarities between the three-state model and the ideas of 
Bouchbinder and Langer, the concepts behind the models are different. For the 
three-state model, the states are inspired by inherent structures
\cite{NAKAGAWA}. Consequently they can be viewed as describing configurational
states. The kinetic fast degrees of freedom are absent from the model and
instead replaced by the thermostat that controls the transition between the
configurational states.

\medskip

As the three-state model belongs to a class of systems described by a master
equation studied in Ref.~\cite{PradosJSTAT} one can ask whether our results
could be deduced from the general results derived in this work, and applied to
complex spin systems in \cite{RUIZ-GARCIA}. However, in the general case
studied in Ref.~\cite{PradosJSTAT} the
analysis of the Kovacs effect can only be made in a linear approximation. This
is interesting in theoretical models because general properties of the Kovacs
hump can be obtained, but it is a severe restriction to analyze experimental
data on glasses. In the study of the Ising model of Ref~\cite{RUIZ-GARCIA}
the linear response results are considered as fine up to temperature jumps
such that the relative change of the relaxation time between the initial
and final temperatures is about ten percent. In glass transitions relaxation
times often change by many orders of magnitudes. In the results presented in
Fig.~\ref{fig:kovacs1}, the relaxation times during aging at $T=0.02$ are 
$\tau_1 = 0.2681\,10^6$ and $\tau_2=0.8708\,10^{12}$
while their values drop to $\tau_1 = 2.471$ and $\tau_2=11.59$
after the temperature jump to
$T=0.17856$ which shows the Kovacs hump. The change is so large
that only exact results, 
which can be obtained for the three-state model, can quantitatively
describe the Kovacs hump in this case.

\medskip
In spite of its simplicity, this model has limitations in the
analytical analysis that it allows. The calculations are simple for all
processes occurring at constant temperature, including out-of-equilibrium
processes triggered by a sharp temperature jump. But for processes involving a
continuous driving of the temperature, the analytical
calculations become unpractical
because they require an expansion on a basis which is continuously evolving.

\medskip
While a thermodynamic definition of the fictive temperature is possible for a
two-level system, it is meaningless for a three-level system. However, it is
possible to define two particular fictive temperatures, each one being
associated to an eigenmode of the system. This suggests a generalization of
the idea of fictive temperature in the case of a real glass.  As shown in
Fig.~\ref{fig:tfict-kovacs}, each fictive temperatures $T_f^{S,1}$,
$T_f^{S,2}$ determines the variation of the fictive temperature
$T_f^{S,\mathrm{transfo}}$ of the system along its trajectory in the
configuration space for a time range around the corresponding relaxation times
$\tau_1$ and $\tau_2$. This suggests that a complex system, with many
relaxation times should be characterized by a distribution of fictive
temperatures. The full distribution is certainly hard to determine, but beyond
the mean which might correspond to the usual concept of fictive temperature,
further moments of this distribution could provide a complementary description
of the thermodynamic state of a complex out-of-equilibrium system.

\medskip
All this approach could be generalized to $N$-level systems, for which $N-1$
independent internal variables are sufficient to follow the non-equilibrium
nature of complex systems. However, moving from $N=2$ to $N=3$ is the
essential step to get a rich behavior because it is sufficient to generate a
system which is not fully determined by its thermodynamic variables only.
The three-state system is the simplest of complex systems!

\appendix
\section{Analytical solution for the time dependence of the occupation
  probabilities}
\label{app-solution}

In terms of the deviations $Q_i$ with respect to equilibrium probabilities, the
two coupled equations for $P_1$ and $P_2$ become an equation for the
vector $\vec{Q}$, having the components $Q_1$ and $Q_2$ which can be written as
\begin{align}
  \label{eq:systemeq}
  \frac{d}{dt} \left( \begin{array}{c} Q_1 \\ Q_2 
                      \end{array} \right)
  =  \left( \begin{array}{cc}
              -A & B \\
              B' & -A' \\
            \end{array} \right)
  \left( \begin{array}{c} Q_1 \\ Q_2 
         \end{array} \right) = M \left( \begin{array}{c} Q_1 \\ Q_2 
         \end{array} \right)
\end{align}
with
\begin{align}
  & A = e^{-B_{12}/T} + e^{-B_{13}/T} + e^{-B_{31}/T} \nonumber \\
  & B = e^{-B_{21}/T} - e^{-B_{31}/T} \nonumber \\
  & A' = e^{-B_{21}/T} + e^{-B_{23}/T} + e^{-B_{32}/T} \nonumber \\
  & B' = e^{-B_{12}/T} - e^{-B_{32}/T} \nonumber \\
\end{align}
To solve equation  (\ref{eq:systemeq}) we can expand $\vec{Q}$ on the
eigenvectors $\vec{U}^{(1)}$ and $\vec{U}^{(2)}$ which diagonalize
the matrix $M$
\begin{align}
  \label{eq:defvectpr}
  M \vec{U}^{(i)} = \lambda_i \vec{U}^{(i)} \; .
\end{align}
The eigenvalues  $\lambda_1$ and $\lambda_2$ are
\begin{equation}
  \label{eq:lambda}
  \lambda_{1,2} = \frac{1}{2} \left[ -(A + A') \pm \sqrt{\Delta} \right]
\end{equation}
with $\Delta = (A - A')^2 + 4 B B'$.
The system parameters which are compatible with the existence of a thermal
equilibrium are such that $\lambda_{1,2} < 0$. Each eigenvalue corresponds to
an eigenvector $\vec{U}^{(i)}$ ($i = 1,2$). Its components are denoted as
\begin{align}
  \label{eq:composantesvectpr}
 \vec{U}^{(i)} = \left( \begin{array}{c} U_1^{(i)} \\ U_2^{(i)} 
                      \end{array} \right)
\end{align}

Matrix $M$ is not a symmetric matrix. It is not orthogonal and it is easy to
check that its eigenvectors are not orthogonal to each other, i.e.
\begin{equation}
  \label{eq:nonortho}
  U_1^{(1)} \; U_1^{(2)} + U_2^{(1)} \; U_2^{(2)} \not= 0 \; .
\end{equation}
However those vector are not colinear
\begin{equation}
  \label{eq:noncolin}
  U_1^{(1)} \; U_2^{(2)} - U_2^{(1)} \; U_1^{(2)} \not= 0
\end{equation}
and therefore they nevertheless define a basis for the $Q_1$, $Q_2$ space.  On
this basis $\vec{Q}$ can be written as
  \begin{equation}
    \label{eq:decomposQ-app}
    \vec{Q} = a(t) \vec{U}^{(1)} + b(t) \vec{U}^{(2)} \; .
  \end{equation}

This calculation holds for any value of the temperature, which could be
time-dependent, as for instance when the system is slowly cooled at a given
rate. However, in practice, the calculation is only useful for all situations
in which the temperature is kept fixed because then the matrix $M$ is
time-independent and so are its eigenvectors. In the following we assume that
the temperature is either a constant or that it evolves by abrupt jumps so
that the temperature protocol can be decomposed in segments during which the
temperature stays constant, as for instance in the protocol used to observe
the Kovacs effect.

Equation  (\ref{eq:decomposQ-app})  defines a system of two scalar equations for
$a$ and $b$. Its determinant is
\begin{align}
  \label{eq:determinant}
  D = \left| \begin{array}{cc} U_1^{(1)} & U_1^{(2)} \\
                               U_2^{(1)} & U_2^{(2)}
                      \end{array} \right| \; .
\end{align}
It does not vanish due to the relation (\ref{eq:noncolin}) so that, if
$Q_1(t)$, $Q_2(t)$ are known, for instance from numerical integration results,
the 
contributions  $a(t)$, $b(t)$ of the two eigenmodes can be determined.

\smallskip
However it is more interesting to get $a(t)$, $b(t)$ by analytically solving
Eq.~(\ref{eq:systemeq}), which leads to
\begin{align}
  \label{eq:systemab}
  \frac {da(t)}{dt} \vec{U}^{(1)} + \frac{db(t)}{dt} \vec{U}^{(2)}
  = \lambda_1 a(t) \vec{U}^{(1)} + \lambda_2 b(t) \vec{U}^{(2)} \;
\end{align}
which can be viewed as a system of two equations for the unknowns
\begin{equation}
  \label{eq:defXY}
  X = \frac {da(t)}{dt} - \lambda_1 a(t)
  \qquad Y = \frac {db(t)}{dt} - \lambda_2 b(t) \; ,
\end{equation}
which can be written
\begin{align}
  \label{eq:systemeXY}
  U_1^{(1)} X + U_1^{(2)} Y &= 0 \nonumber \\
  U_2^{(1)} X + U_2^{(2)} Y &= 0  \; .
\end{align}
The determinant of this system is again the determinant $D$ of
Eq.~(\ref{eq:determinant}), which is non-zero. As the right-hand-side of the
system is zero, the only solution of the system is $X=0$, $Y=0$. According to
(\ref{eq:defXY}) it implies that the general solutions for $a(t)$ and $b(t)$
are exponential relaxations
\begin{align}
  \label{eq:solab-app}
  a(t) &= a(t=0) \exp[ - t / \tau_1] \nonumber \\
  b(t) &= b(t=0) \exp[ - t / \tau_2]
\end{align}
where
$\tau_{1,2} = -1/\lambda_{1,2}$.

\end{document}